\documentclass[prb,amsfonts,amssymb,floats,twocolumn,superscriptaddress,aps]{revtex4-1}
\usepackage[T1]{fontenc}
\usepackage{amsmath}
\usepackage{float}
\usepackage{color}
\usepackage{braket}
\usepackage{bm}
\usepackage{dsfont}
\usepackage{adjustbox}
\usepackage{graphicx}
\usepackage{enumitem}
\usepackage{bbm}
\usepackage{xtab,afterpage}
\usepackage{longtable}
\usepackage{scrextend}

\usepackage{array}
\usepackage{adjustbox}

\newcolumntype{M}[1]{>{\centering\arraybackslash}m{#1}}
\newcommand\marginv{3}
\newcommand\scalef{0.4}

\usepackage[colorlinks=true]{hyperref}
\hypersetup{linkcolor=blue,citecolor=blue,urlcolor=blue}

\usepackage[normalem]{ulem}

\newcommand{\iu}{\mathrm{i}} 
\newcommand{\eu}{\mathrm{e}} 
\newcommand{\du}{\mathrm{d}} 
\newcommand{\hc}{\mathrm{h.c.}} 

\newcommand{\Uone}{\mathrm{U(1)}}
\newcommand{\SUtwo}{\mathrm{SU(2)}}

\newcommand{\SO}{\mathrm{SO}}

\newcommand{\umat}[1]{{\underline{\underline{#1}}}}
\newcommand{\uvec}[1]{{\underline{#1}}}
\newcommand{\bvec}[1]{{\bm{#1}}}

\DeclareMathOperator{\tr}{tr}
\DeclareMathOperator{\diag}{diag}

\DeclareMathOperator{\im}{Im}

\graphicspath{{./}{figs/}}

\begin{document}

\title{Phase diagrams and excitations of anisotropic $S=1$ quantum magnets\\on the triangular lattice}

\author{Urban F. P. Seifert}
\affiliation{Univ Lyon, ENS de Lyon, CNRS, Laboratoire de Physique, 69342 Lyon, France}
\affiliation{Kavli Institute for Theoretical Physics, University of California, Santa Barbara, CA 93106}
\author{Lucile Savary}
\affiliation{Univ Lyon, ENS de Lyon, CNRS, Laboratoire de Physique, 69342 Lyon, France}


\date{\today}

\begin{abstract} The $S=1$ bilinear-biquadratic Heisenberg exchange model on the triangular lattice with a single-ion anisotropy has previously been shown to host a number of exotic magnetic and nematic orders [Moreno-Cardoner \textit{et al.}, Phys.~Rev.~B \textbf{90}, 144409 (2014)], including an extensive region of ``supersolid'' order. In this work, we amend the model by an XXZ anisotropy in the exchange interactions. Tuning to the limit of an exactly solvable $S=1$ generalized Ising-/Blume-Capel-type model provides a controlled limit to access phases at finite transverse exchange. Notably, we find an additional macroscopically degenerate region in the phase diagram and study its fate under perturbation theory. We further map out phase diagrams as a function of the XXZ anisotropy parameter, ratio of bilinear and biquadratic interactions and single-ion anisotropy, and compute corrections to the total ordered moment in various phases using systematically constructed linear flavor-wave theory. We also present linear flavor-wave spectra of various states, finding that the lowest-energy band in three-sublattice generalized (i.e.\ with $S^z=\pm1,0$) Ising/Blume-Capel states, stabilized by strong exchange anisotropies, is remarkably flat, opening up the way to flat-band engineering of magnetic excitation via stabilizing non-trivial Ising-ordered ground states. 
\end{abstract}

\maketitle


\section{Introduction}

\subsection{Motivation and model}

The antiferromagnetic Heisenberg model on the triangular lattice is a paradigmatic model for local moments experiencing frustrated interactions.
While some exchange interactions (such as Heisenberg nearest-neighbor interactions) are not sufficient to drive the system into a quantum-disordered spin-liquid phase on this lattice, even for the case of $S=1/2$ moments, the frustrated character manifests itself in unconventional ordered states, in particular non-collinear and non-coplanar states. Typically, those experience sizeable quantum fluctuations, and the order parameter amplitude is reduced compared with the classical case.
Notably, considering $S=1/2$ moments, moving away from the Heisenberg limit and tuning the relative strength of transverse and longitudinal nearest neighbor interactions (i.e.\ considering a XXZ-type Hamiltonian) gives rise to a rich phase diagram featuring intriguing ``supersolid'' phases characterized by simultaneous superfluid (in-plane $\Uone$-symmetry breaking) and crystalline (longitudinal $\mathbb{Z}_2$-symmetry breaking) order \cite{melko05,wang09}.

The aforementioned unconventional phases feature long-range order of ``dipolar'' local moments, the only possible type of ``spin order'' in the case of $S=1/2$.
In recent years, ordered states in quantum magnets with higher spin, for example $S=1$ systems such as NiGa$_2$S$_4$ \cite{naka05}, Ba$_3$NiSb$_2$O$_9$ \cite{chen10} or the $S=3/2$ Ba$_2$CoGe$_2$O$_7$ \cite{romh12}, as well as two-dimensional van-der-Waals materials \cite{novo19,kart20}, have received increased experimental and theoretical attention \cite{tsunetsugu2006,laeuch06,PhysRevB.79.214436,penc11,penc12,bai21,corboz2007spontaneous,xu07,niesen2017tensor,niesen2018ground}.
In contrast to the spin-$1/2$ case where no such terms exist, multipolar interactions and single-ion anisotropies may then be sizeable and stabilize ordered states in which the order parameter transforms in some higher-dimensional representation of $\SUtwo$ (note that ``mixed'' dipolar and multipolar orders are also conceivable).
These long-range ordered phases are intrinsically contingent on the quantum nature of spin in the sense that they do not have a classical counterpart in the conventional ``classical'' limit of $S\to\infty$ commonly studied for dipolar-ordered states (we provide an alternate notion of a ``classical'' limit applicable also for multipolar states below). Like their dipolar counterparts, multipolar terms may also be ``frustrated'' and anisotropic, and, in addition to ``collinear'' multipolar ordered states, yet more exotic order parameter patterns can emerge on frustrated lattices, such as the triangular one \cite{penc11,tothesis}.
Experimentally, the multipolar nature of magnetic excitations of spin-1 moments (corresponding to a $J_\mathrm{eff}=1$ manifold emerging from the spin-orbit-split $t_{2g}$ levels) on the triangular lattice in FeI$_3$ has been the focus of the recent work in Ref.~\onlinecite{bai21}.

Motivated by these recent developments, we reconsider the spin-1 bilinear-biquadratic model on the triangular lattice \cite{tsunetsugu2006,laeuch06} with single-ion anisotropy previously discussed\cite{moreno14}, and further consider XXZ exchange anisotropy, both at the bilinear and biquadratic levels, namely we study the following Hamiltonian for $S=1$
\begin{align} \label{eq:h0}
	\hat{\mathcal{H}} &= \hat{\mathcal{H}}_1 + \hat{\mathcal{H}}_2 + \hat{\mathcal{H}}_D \nonumber\\ &= J_1 \sum_{\langle ij \rangle} \left[\lambda \left( \hat{S}^x_i \hat{S}^x_j + \hat{S}^y_i \hat{S}^y_j \right) + \hat{S}^z_i \hat{S}^z_j \right] \nonumber\\
	&\quad+ J_2 \sum_{\langle ij \rangle} \left[\lambda \left( \hat{S}^x_i \hat{S}^x_j + \hat{S}^y_i \hat{S}^y_j \right) + \hat{S}^z_i \hat{S}^z_j \right]^2 + D \sum_i \left(\hat{S}^z_i\right)^2.
\end{align}
Eq.~\eqref{eq:h0} has the advantage of capturing what we expect to be the main source of anisotropy in $S=1$ anisotropic triangular materials, i.e.\ that where in-plane isotropy remains, and yet has a manageable number of parameters while still containing several limits of interest.

\begin{figure*}[htbp]
  \centering
  \includegraphics[width=\textwidth]{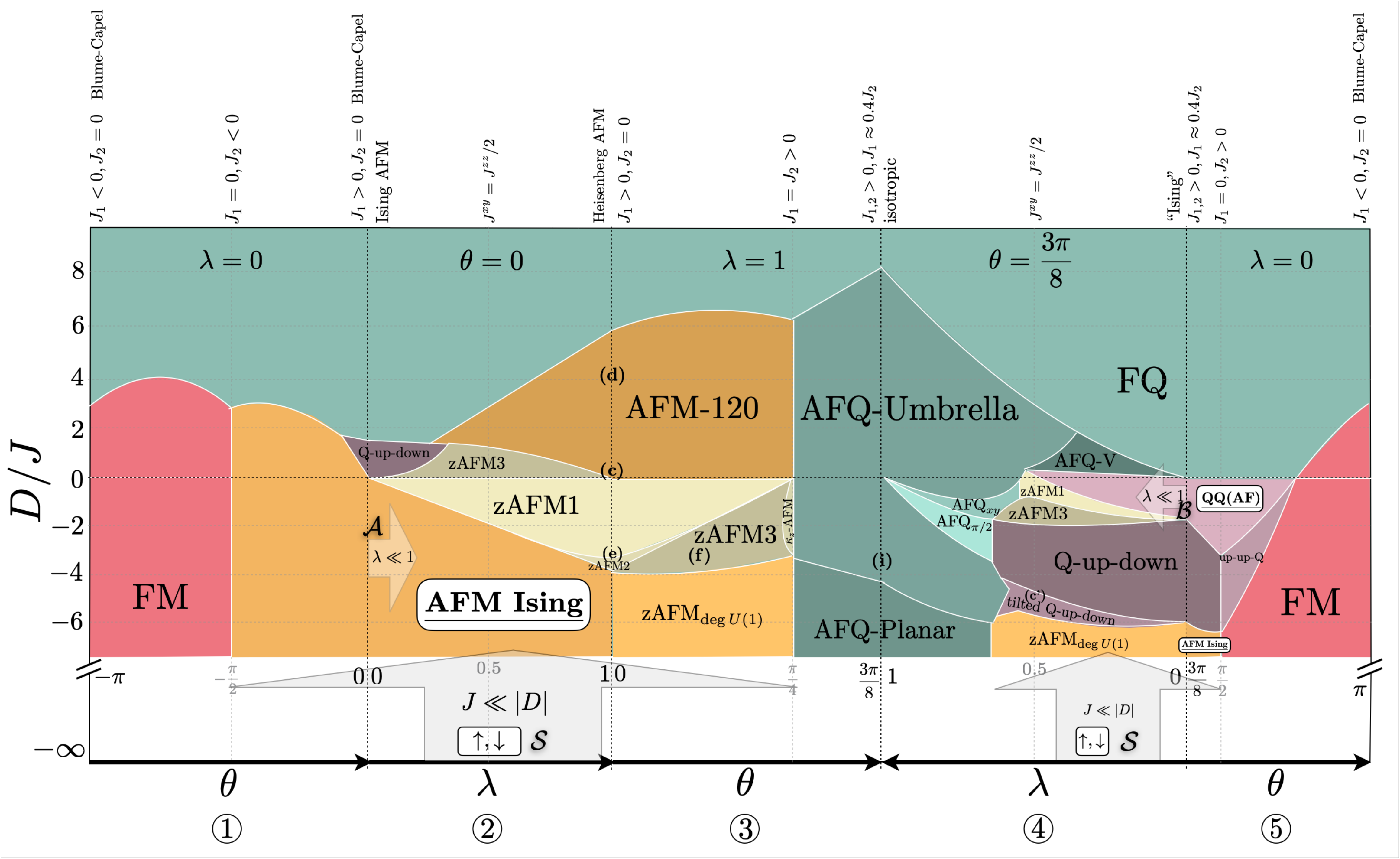}
  \caption{Summary of the phase diagrams derived in this manuscript from Eq.~\eqref{eq:h0} through a variational approach using three-sublattice product states. The circled numbers correspond to path segments shown in Fig.~\ref{fig:side}(a). For phase diagrams shown with color plots of the ``moment'' reduction, see Figs.~\ref{fig:pd_theta_D} ($\lambda=1$), \ref{fig:pd_ising} ($\lambda=0$, exact), \ref{fig:lambdaCut_0} ($\theta=0$), \ref{fig:lambdaCut_3pi8} ($\theta=3\pi/8$). All phases are described in Table~\ref{tab:states}. The phases whose name is shown in a white box are macroscopically degenerate within the classical approximation. Perturbation theory near regions $\mathcal{A}$, $\mathcal{S}$ and $\mathcal{B}$ is described in Sections~\ref{sec:lambdall1}, \ref{sec:pert-JDll1} and \ref{sec:degen-3pi8}, respectively. The (c,d,e,f,i) labels are shown at the location in parameter space of the excitation spectra displayed in the corresponding panels in Fig.~\ref{fig:dssLambdaOne}, while (c') corresponds to panel (c) in Fig.~\ref{fig:dssLambdaSmall}.}
  \label{fig:alltogether}
\end{figure*}

Indeed, our model captures: {\em (i)} $\lambda=1,D=0,J_2=0$, which is the triangular Heisenberg model, which orders both classically (and for $S=1/2$) into a three-sublattice (``120 degree'') antiferromagnet \cite{zheng06,caprio99,bernu92}, while {\em (ii)} $\lambda=0,D=0,J_2=0$ is the $S=1$ triangular Ising antiferromagnet, whose $S=1/2$ counterpart has been extensively studied in the literature and is known to have a macroscopic ground-state degeneracy at zero temperature $T=0$ \cite{wannier50,moes01,steph64}, {\em (iii)} the $|D|\gg|J_1|,|J_2|$ quantum model with $D<0$ which reduces to an effective $S=1/2$ model.
It is interesting to understand how far such phases extend away from these limiting cases, and what the neighboring phases are.
Moreover, an analysis away from tractable limits can in fact provide insight into the nature and stability of the ground states in the limiting cases, should, e.g., a given phase continuously connect several limits.
Other models included in our Hamiltonian, such as the $D=0,\lambda=1$ (quantum) model were studied in Ref.~\onlinecite{laeuch06} by means of exact diagonalization and linear flavor-wave theory (see also Ref.~\onlinecite{tsunetsugu2006}) and using tensor networks in Ref.~\onlinecite{niesen2018ground}, the $\lambda=1$ model has been analysed using mean-field theory by T\'oth \cite{tothesis} and in Ref.~\onlinecite{moreno14} using cluster mean-field theory.
Further, the XXZ model on the triangular lattice (i.e.\ $D=0=J_2$) with \emph{classical} spins was studied in Refs.~\onlinecite{kawa85,henley92,kleine92}.
Murthy, Arovas and Auerbach performed a semiclassical analysis \cite{murthy97}, and more recent numerical studies \cite{melko05,wang09,jiang09} have provided insight into the quantum $S=1/2$ model. Finally, we note that $S=1$ models can be obtained in the large Hund limit of some spin-orbital models, such as SU(4) ones which have been for example studied in the context of twisted bilayer graphene.\cite{keselman2020,kiese2020}

\subsection{``Classical'' order in multipolar spin systems} \label{sec:intro-concept}

Let us now turn to the physical interpretations of some of the ground states of the model above,
for which we seek to develop a description in terms of local mean fields, with ground-state configurations given by the set of fields that minimize the mean energy $E = \langle \mathcal{H} \rangle$ for a given set of parameters.
Considering magnetic systems, such a mean-field description is often understood to correspond to taking a ``classical'' limit wherein the $\SUtwo$ spin operators $\hat{\mathbf{S}}$ at each site are replaced by scalar vectors $\vec{\mathsf{S}}$ through $\hat{\mathbf{S}}\rightarrow\vec{\mathsf{S}}=\langle\psi|\hat{\mathbf{S}}|\psi\rangle$.
If one is interested in systems with dipolar order, the state $\ket{\psi} = \ket{S,\phi,\theta}$ can be taken to be a Bloch coherent spin state such that $\vec{\mathsf{S}}$ is a three-dimensional unit vector of length $S$ and polar angles $\theta$ and $\phi$. In the limit of $S\to\infty$, one further has \emph{at leading order} $\langle \hat{S}^\alpha \hat{S}^\beta \rangle = \mathsf{S}^\alpha \mathsf{S}^{\beta} + \mathcal{O}(S)$ \cite{lieb73}, justifying the folklore notion of ``replacing spin operators by classical vectors''.
This approach, while acceptable in dipolar-ordered phases, comes short of describing the nematic phases accessible with $S=1$ spins (or any $S > 1/2$), and which appear for example in the $D=0,\lambda=1$ phase diagram, driven by a nonzero $J_2$, as discussed in Ref.~\onlinecite{laeuch06}.
Instead, one makes progress by introducing mean fields for each hermitian operator acting non-trivially on the Hilbert space of spin-$S$.
For $S=1$, in addition to the spin operators $\hat{S}^\alpha$, these operators are given by five linearly independent components of the (symmetric, traceless) quadrupole operator matrix $\hat{Q}^{\alpha \beta} = \hat{S}^\alpha \hat{S}^\beta + \hat{S}^\beta \hat{S}^\alpha - 4/3 \delta^{\alpha\beta} \mathds{1}$, for which one can introduce mean fields $\mathsf{Q}^{\alpha \beta} \equiv \braket{\psi | \hat{Q}^{\alpha \beta} | \psi}$ in any state $\ket{\psi}$ in the $S=1$ Hilbert space.
While this leads to a total of $3+5=8$ mean fields, we note that they cannot be chosen independently.
Instead, as discussed below, the expectation values of all hermitian operators acting on a $S=1$ moment are fully specified in terms of 4 independent parameters which can be optimised using variational searches. From now on, we will refer to the model defined by the energy \emph{function} $E=\langle \hat{\mathcal{H}} \rangle$, expressed in terms of $\vec{\mathsf{S}}$ and $ \mathsf{Q}^{\alpha \beta}$, as the ``classical'' version of \eqref{eq:h0}, as it is defined on a configuration space (rather than a Hilbert space) and no longer contains quantum fluctuations.

As an illustrative example of the discussion above, we discuss the role and interpretation of the single ion anisotropy $\mathcal{H}_\mathrm{SI} = D (\hat{S}^z)^2$. 
In the quantum case, $D$ splits the $S^z=0,\pm1$ triplet into a $S^z=\pm1$ doublet and a $S^z=0$ singlet.
In the ``dipolar'' classical limit, $D$ favors the spins to be in-plane ($D>0$), corresponding to a 1-parameter family of states, or along the easy $z$ axis ($D<0$), as seen from $\langle \mathcal{H}_\mathrm{SI} \rangle = S^2  D \cos^2 \theta$.
In the generalized model introduced above, $D>0$ is seen to favor $\mathsf{Q}^{zz}= -4/3$ with a unique ground state (in contrast to the dipolar limit), while for $D<0$, the ground state which has $\mathsf{Q}^{zz}=2/3$ is twofold degenerate (spontaneously breaking the in-plane rotational symmetry), so that the ground-state manifold can be parametrised in terms of two continuous parameters (equivalent to the two-level Bloch sphere), therefore accurately reflecting the two-fold degeneracy of the aforementioned doublet. 
Note that the singlet state $\ket{0}$ has $\braket{0| S^\alpha | 0} \equiv 0 \ \forall \alpha$ and thus cannot be represented in terms of a coherent spin state (which in the limit $S\to\infty$ yield ``classical'' dipolar spin states).
However, from $\braket{0| \left(S^x\right)^2 | 0} = \braket{0| \left(S^y\right)^2 | 0} = 1$ and $\braket{0| \left(S^z\right)^2 | 0} = 0$ it becomes clear that spin fluctuations in the $S^z=0$ state are no longer isotropic, but rather orthogonal to the $\hat{z}$-axis, paving the way towards the definition of a ``director'' as an order parameter for nematic orders, which is orthogonal to the plane of spin fluctuations.

\subsection{Summary of results and outline}


In light of the above comments, we derive the {\em three-sublattice} semiclassical phase diagram of the model Eq.~\eqref{eq:h0}: This matches the natural ``magnetic'' unit cell for the triangular lattice which enlarges the structural one three times and indeed has been found relevant in a large region of the phase diagram of the isotropic $S=1$ model.\cite{laeuch06,moreno14} In order to do so, we use a variational approach and calculate the quantum fluctuations throughout the phase diagram (Secs.~\ref{sec:moment-reduc} and \ref{sec:disc-phase-diagr}). We find in particular that those are the largest near most phase transitions, in regions where several phases compete, and particularly so in the region where ``supersolid phases'' arise (Fig.~\ref{fig:pd_theta_D}, \ref{fig:lambdaCut_0} and \ref{fig:lambdaCut_3pi8}). We further make use of our flavor wave theory to compute the dynamical structure factor as obtained through neutron scattering (Sec.~\ref{sec:spectr}).

\begin{figure}[htbp]
  \centering
  \includegraphics[width=\columnwidth]{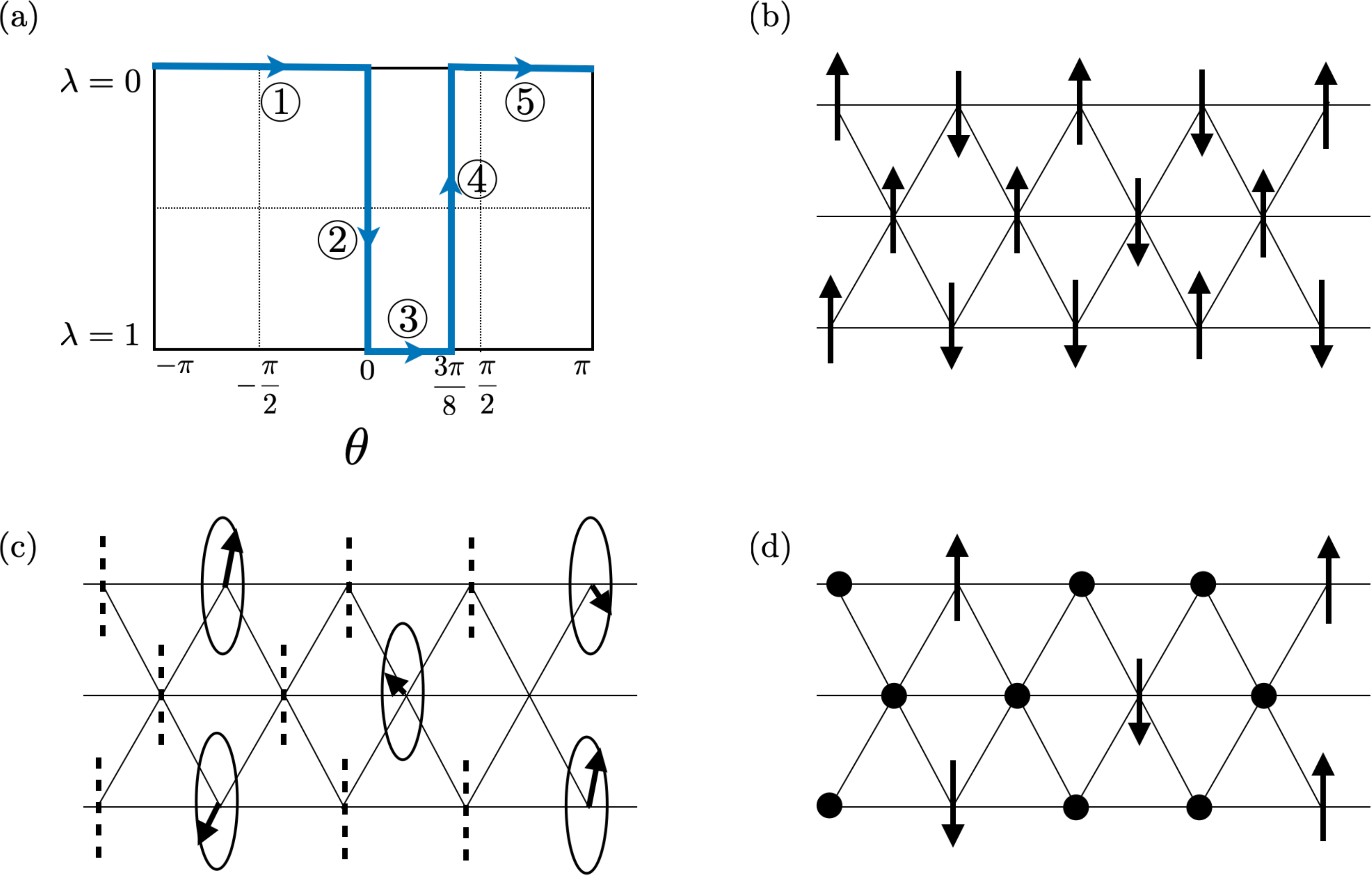}
  \caption{(a) Path in parameter space for the phase diagram shown in Fig.~\ref{fig:alltogether}. (b) A given realization of the macroscopically degenerate manifold of the Ising AFM phase. (c) Classical picture of a given realization of a state in the QQ(AF) phase. The dashed line represents a director in the $z$ direction. The circle with an arrow is meant to illustrate a possible classical dipolar state in the degenerate $U(1)$ manifold. (d) Quantum picture of a given realization of a state in the QQ(AF) phase. The black dot denotes the $|S^z=0\rangle$ quantum state.}
  \label{fig:side}
\end{figure}

Another central result of our work is the particularly rich phase diagram driven by $\lambda\neq1$ (see especially Figs.~\ref{fig:lambdaCut_3pi8} and \ref{fig:alltogether}). We find, in addition to the phases which exist in the $J_1$-$J_2$-$D$ phase space, that $\lambda\neq1$ stabilizes some phases whose magnetic unit cell comprises both quadrupolar and dipolar states on different sites, namely an ``up-down-quadrupolar,'' a tilted version thereof, an ``up-up-quadrupolar,'' and a ``quadrupolar-quadrupolar-AF.''
The latter can be seen to extend from the $\lambda=0$, $D=0$ for $\pi/4\leq\theta\leq3\pi/4$ segment to a large region of the phase diagram at $D<0$ (Fig.~\ref{fig:pd_ising}) and up to $\lambda=0.6$ (for $\theta=3\pi/8$, Fig.~\ref{fig:lambdaCut_3pi8}).
In its quantum mechanical version, the spin is in its zero state in two of the three sublattices, while the third is free, {\em independently at every such site}, to be in a $S^z=+1$ or $S^z=-1$ state (see Fig.~\ref{fig:side}(c,d) for an illustration).
In that sense, this state is a $1/3$ sublattice version of the Ising antiferromagnet on the triangular lattice, and is macroscopically degenerate.
While exact in the limit of purely longitudinal (local $S^z$-conserving) interactions, for finite transverse exchange this degeneracy is an artefact of the mean-field approach in which every third (dipolar) spin experiences zero net exchange interactions.
In turn, in order to investigate the real ground state, we do promote the model to a quantum one and perform perturbation theory in $\lambda\ll1$ around the $\lambda=0$ manifold (``region $\mathcal{B}$'', Sec.~\ref{sec:degen-3pi8}), i.e.\ in the limit of Ising interactions, which is exact, even quantum mechanically.
Intermediate sites with $S^z=0$, and the existence of three spin states at each site lead to a very different perturbation theory and in turn to very different behavior from that of the conventional all-sites, $S=1/2$, Ising antiferromagnet.
Indeed, we predict that plaquette terms which arise at second order in perturbation theory, may lead to a quantum spin liquid when the bilinear part of the exchange is antiferromagnetic.
See Table~\ref{tab:degpert} and Sec.~\ref{sec:degen-3pi8}.

Finally, we show that the variational treatment predicts the stability of the $S=1$ Ising antiferromagnet all the way to the $|D|\gg J$ ($D<0$)---``region $\mathcal{S}$''---and $\lambda=0$---``region $\mathcal{A}$''---limits, and that perturbation theory from both limits (see Sec.~\ref{sec:region-A}) {\em produces} effective Hamiltonians which are known to give rise to two supersolid phases,\cite{melko05,moreno14} therefore showing that these limits are connected and confirming that the supersolid phases are stable in a large region of our phase diagram. 

The remainder of the paper is structured as follows.
We introduce the model and present further details on the variational ground state search, as well as a pedagogical presentation of linear flavor-wave theory and computed order parameter amplitude corrections in Sec.~\ref{sec:model-LFWT}. The resulting phase diagrams are shown and discussed in Sec.~\ref{sec:disc-phase-diagr}.
In Sec.~\ref{sec:exact-Ising-lim}, we employ perturbation theory and mapping to models in the literature to discuss the lifting of macroscopic ground-state degeneracies and access phases beyond mean-field theory.
We show exemplary linear-flavor wave spectra and dynamical spin structure factors in Sec.~\ref{sec:spectr}, before we close the paper with a discussion in Sec.~\ref{sec:conclusio}.

\begin{table}[htbp]
  \centering
  \begin{tabular}{cccc}
    \hline\hline
    $|D/J|=\infty$ & decoupled $S_i^{\rm eff}=1/2$\\
    \hline
    $|D/J|\ll1$ & $S^{\rm eff}=1/2$, XXZ model & $\theta<0$ &  supersolid 1\\
    && $\theta>0$ & supersolid 2\\
    \hline\hline
    $\lambda=0$ & decoupled $S_{i\in{\rm C}}^{\rm eff}=1/2$\\
    \hline
    $\lambda\ll1$ & $S^{\rm eff}\!=\!1/2$, Ising+plaquette &$\theta<\pi/2$ & QSL? \\
    &&$\theta>\pi/2$ & $|\sum_i S_i^z|=1/3$ \\
    \hline\hline
  \end{tabular}
  \caption{Phases obtained from degenerate perturbation theory from the manifold spanned by $\{|S^z_i=1\rangle,|S^z_i=-1\rangle\}$ ``region $\mathcal{S}$'' (top), and that spanned by ``region $\mathcal{B}$'' $\{|0,0,S^z_{i\in{\rm C}}=1\rangle, |0,0,S^z_{i\in{\rm C}}=-1\rangle\}$ (bottom). ``supersolid 1'' is characterized by an effective $S=1/2$ in-plane expectation value $\langle\sigma^+\rangle=m^\perp (1,1,1)$ on the three sublattices. ``supersolid 2'' is characterized by $\langle\sigma^+\rangle=m^\perp (1,-1,0)$ and $\langle\sigma^z\rangle= (m^z,-m^z,2m^z-\delta)$.}
  \label{tab:degpert}
\end{table}

\section{Model, classical ground states and flavor-wave theory} \label{sec:model-LFWT}

\subsection{Hamiltonian}

The biquadratic part of the Hamiltonian can be rewritten in a basis of (linearly independent) spin and \emph{quadrupolar} operators $\hat{Q}^{\alpha \beta} =  \hat{S}^\alpha \hat{S}^\beta + \hat{S}^\beta \hat{S}^\alpha -4/3 \delta^{\alpha \beta} \mathds{1}$ (which are traceless and symmetric in $\alpha,\beta$, and we drop the hat symbol for operators from now on), yielding (we henceforth shall drop the hat symbol for operators) 
\begin{align}
	\mathcal{H} = &\sum_{\langle ij \rangle} \Bigg[ \tilde{J}_1^{xy}\left(S^x_i S^x_j + S^y_i S^y_j\right) + \tilde{J}_1^{z}S^z_i S^z_j + \frac{J_2}{4} \tr\left[ \Lambda Q_i \Lambda Q_j \right] \Bigg] \nonumber\\
	&+\sum_i \left[ \tilde{D}\left(S^z_i \right)^2 \right],
\end{align}
with $\tilde{J}_1^{xy}=\lambda\left(J_1- \frac{J_2}{2}\right)$, $\tilde{J}_1^{z}=\left(J_1-\frac{J_2 \lambda^2}{2}\right)$, $\tilde{D}=\left(D + 4(1-\lambda^2) J_2 \right)$,
and where $\Lambda = \diag(\lambda,\lambda,1)$ and the trace is taken over the $\SUtwo$ algebra indices. We have dropped a global additive constant. As in other works, we parametrize the bilinear and biquadratic exchange couplings as $J_1 = J \cos \theta$ and $J_2 = J \sin \theta$ ($J>0$). In particular $\theta=0,\pi$ contains only bilinear terms, $J_2=0$, while $\theta=\pm\pi/2,\pm3\pi/2$ have $J_1=0$.
 
As mentioned above, we start with the determination of the phase diagram obtained semiclassically, for which we estimate the quantum fluctuations. 

 \subsection{Parametrization of ground states} 

We first perform a mean-field approximation to find the ground states of the Hamiltonian \eqref{eq:h0} in a variational manner.
To fully parametrize all states in the Hilbert space of a $S=1$ moment, it is convenient to choose a basis in which $S^\alpha \ket{\alpha} = 0$ (no summation) for $\alpha = x,y,z$, which is achieved by taking $\ket{x} = \iu \left(\ket{+1} - \ket{-1}\right)/\sqrt{2}$, $\ket{y} = \left(\ket{+1}+\ket{-1}\right)/\sqrt{2}$ and $\ket{z} = - \iu \ket{0}$ \cite{penc11}.
Any state in the $S=1$ Hilbert space can then be written as $\ket{\uvec d} = \sum_\alpha d^\alpha \ket{\alpha}$ where $\uvec d = \uvec u + \iu \uvec v$ with $\uvec{u},\uvec{v} \in \mathbb{R}^3$ is a complex three-dimensional vector which is normalized $|\uvec d|^2 =1 = |\uvec u|^2 + |\uvec v|^2$.
Here, we will use underlines to denote specifically elements in the ``director space'' $\uvec{d} \in \mathbb{C}^3$ and use more generally arrows $\vec{\cdot}$ for objects transforming under $\SO(3)$ spin rotations.

In the above basis, the spin operators are written as $S^\alpha = -\iu \sum_{\beta,\gamma} \epsilon^{\alpha \beta \gamma} \ket{\beta}\bra{\gamma}$.
The expectation value in the state $\ket{d}$ of a spin operator and a spin bilinear are easily evaluated as
\begin{equation} \label{eq:s_dvec}
	\braket{\vec S} = -\iu \uvec d^\ast \times \uvec d \quad \braket{S^\alpha S^\beta} = \delta^{\alpha \beta} - (d^\ast)^\alpha d^\beta.
\end{equation}
We see that states described by vectors $\uvec d$ which are real up to a global phase have vanishing $\braket{\vec S} = 0$.
A physical observable for these \emph{quadrupolar} states is given by the symmetric, traceless quadrupole operators $Q^{\alpha \beta}$ which have five independent components.
In the $\uvec d$-formalism, the expectation value of a local quadrupolar operator takes the form
\begin{equation} \label{eq:q_dvec}
	\langle Q^{\alpha \beta} \rangle = \frac{2}{3}\delta^{\alpha \beta} - (d^\ast)^\alpha d^\beta - d^\alpha (d^\ast)^\beta.
\end{equation}
We note that, given the definition of $\ket{\uvec d}$, there is a $\Uone$ arbitrariness in defining the phase of the $\uvec d$ (rendering it a \emph{director}, rather than a vector) as also visible from the invariance of \eqref{eq:s_dvec} and \eqref{eq:q_dvec} under global phase rotations $\uvec d \mapsto \eu^{\iu \varphi} \uvec d$.

Noting that $\mathbb{C}^3 \simeq \mathbb{R}^6$, we can parametrize the normalized vector $(\uvec u, \uvec v )^\top \in \mathbb{R}^6$ in terms of four angles (note that we have the freedom to fix the global phase of $\ket{d}$, e.g.\ by demanding $\im[d^z] \equiv 0$) as
\begin{equation} \label{eq:dvec-angles}
	\uvec d = \begin{pmatrix}
		\cos \varphi_1 + \iu \sin \varphi_1 \sin\varphi_2 \sin \varphi_3 \cos \varphi_4 \\
		\sin \varphi_1 \cos \varphi_2 + \iu \sin \varphi_1 \sin\varphi_2 \sin \varphi_3 \sin \varphi_4 \\
		\sin \varphi_1 \sin \varphi_2 \cos \varphi_3
	\end{pmatrix}.
\end{equation}
For the (analytical) construction of phase boundaries, it is useful to find ``minimal parametrizations'' of the $\uvec{d}$-vector which yield one element of the symmetry-degenerate ground state manifold in a given phase, with as few parameters as possible.
Minimizing the ground state energy with respect to these ``minimal'' parameters then often allows us to find analytical expressions for the ground-state energy in a phase $\mathsf{p}$ as a function of the external control parameters $E_{\mathsf{p}}(\theta,\lambda,D)$ (in more complex ordering patterns, the optimal minimal parameters are defined only through implicit equations which require a numerical solution).
By demanding $E_{\mathsf{p}_i} = E_{\mathsf{p}_j}$ one then obtains an implicit equation for the phase boundary between phases $\mathsf{p}_i$ and $\mathsf{p}_j$ in $\theta$-$\lambda$-$D$-space.
In some cases, we are further able to determine a closed form expressions for the associated curve.

\subsection{Linear flavor-wave theory}

Magnons as excitations of long-ranged ordered magnets can be described in terms of linear spin-wave theory, which formally corresponds to the quadratic piece in a $1/S$ expansion about a classical reference state as a local minimum of the free energy of the system.
Crucially however, the standard Holstein-Primakoff approach involves expanding about a classical \emph{dipolar} reference state.
While this expansion can be evaluated for any value of $S$, this crucially neglects fluctuations of classical reference states which have non-zero multipolar components (or, more drastically, possess only some multipolar order) as they may occur for representations of $\SUtwo$ with $S>1/2$.
These states and their fluctuations thus cannot be accurately described in this approach.

Hence, to compute the spectra of non-dipolar ordered phases we employ linear flavor-wave theory \cite{tsunetsugu2006,laeuch06,penc11,batista14}, which can be understood as a generalized spin-wave theory, by taking advantage of the above parametrization in terms of $\uvec d$ vectors.

\subsubsection{Conceptual overview}

\emph{First}, we introduce a set of Schwinger bosons $b^\alpha$ with three flavors $\alpha=x,y,z$ corresponding to the basis states of our $S=1$ Hilbert space with $b_\alpha^\dagger \ket{0} = \ket{\alpha}$.
Imposing the unit filling constraint $\sum_\alpha b_\alpha^\dagger b_\alpha =1$, it is easily seen that the representation of spin operators
\begin{equation} \label{eq:s_schwinger_bosons}
	S^\alpha = -\iu\sum_{\beta,\gamma} \epsilon^{\alpha \beta \gamma} b_\beta^\dagger b_\gamma
\end{equation}
satisfies the $\SUtwo$ algebra.

\emph{Second}, to describe fluctuations about the ordered state, we condense one of the bosons above, say $b_z$, and perform a generalized Holstein-Primakoff transformation $b_z, b_z^\dagger \to \sqrt{M- b_x^\dagger b_x - b_y^\dagger b_y}$ where we generalize the unit filling constraint to allow for $M$ bosons per site, with $M=1$ corresponding to the physical case.
The virtue of considering a general $M$ is that we can now systematically expand the Hamiltonian as well as all operators of interest in $1/M$, i.e.
\begin{equation}
	\mathcal{H} = M^2 \mathcal{H}^{(0)} + M \mathcal{H}^{(2)} + \mathcal{O}(\sqrt{M}),
\end{equation}
where $\mathcal{H}^{(0)}$ does not contain bosonic operators and $\mathcal{H}^{(2)}$ is quadratic in the bosons. Truncating the above expansion at $\mathcal{H}^{(2)}$ corresponds to the harmonic approximation, yielding \emph{linear} flavor-wave theory.
If the classical reference state corresponds to a local minimum in the free energy, no terms linear in the bosons appear in the above expansion.

In the limit $M\to\infty$, where the leading order terms in $1/M$ dominate, quantum fluctuations are suppressed and the model is reduced to a ``classical'' system (with the notion of classicality as introduced in Sec.~\ref{sec:intro-concept}), the ground state of which can be determined variationally \cite{penc11}.
However, we emphasize that for $\mathcal{H}$ to indeed reproduce the ``classical'' Hamiltonian in the limit of $M \to \infty$ (i.e. for $\mathcal{H}^{(0)}$ to match the mean-field Hamiltonian $\langle \mathcal H \rangle$ obtained from Eq.~\eqref{eq:h0}), we need to rescale $D_\mathrm{eff} \equiv D + 4(1+\lambda^2) J_2 \mapsto M D_\mathrm{eff}$, which is analogous to the rescaling of an external magnetic field in standard spin-wave theory \cite{coletta12}.

To make contact with experiment, for computing spectra and evaluating physical observables, we take $M=1$ after making the aforementioned harmonic approximation.
For a formal discussion of flavor-wave theory as a generalized spin-wave theory we refer the reader to Ref.~\onlinecite{batista14}.

\subsubsection{Procedure}

In practice, the choice of the classical reference state to expand about is determined by the selection of the boson (or the linear combination of bosons) to be condensed, where the (dominant) $\sim\sqrt{M}$-contribution to the bosonic spinor $b_\alpha$ after the condensation determines the $\alpha$-component of the $\uvec{d}$-vector.
This is particularly evident when comparing Eqs.~\eqref{eq:s_dvec} and \eqref{eq:s_schwinger_bosons}.

To construct a flavor-wave theory with a generic $\uvec d$-vector determining the classical reference frame, we therefore first perform a unitary transformation $U$ to a \emph{canonical} basis (primed) so that $\uvec d' = \umat{U} \uvec d \equiv (0,0,1)^\top$, and similarly $\uvec{b}' = \umat{U} \uvec{b}$ where $\uvec{b}=(b_x,b_y,b_z)^\top$, so that the ordered state is obtained by condensing $b_z'$ in the canonical basis.
The matrix $\umat{U}$ can be constructed as
\begin{equation} \label{eq:umat}
	\umat{U}^\dagger = (\uvec u,\uvec v,\uvec d)^\top,
\end{equation} 
where $\uvec u$ and $\uvec v$ are two (arbitrary) vectors so that $\uvec u, \uvec v$ and $\uvec d$ form an orthonormal basis, see also Appendix~\ref{sec:def_trafo} for an explicit construction. 

In this work, we first use symbolic manipulations in \texttt{Mathematica} to derive analytic expressions for the Hamiltonian for a general choice of $\umat{U}_s$ on the $s=A,B,C$ sublattices, condense the $b_z'$ boson and subsequently perform a Fourier transformation to extract the boson bilinear term
\begin{equation} \label{eq:H2biq}
	\mathcal{H}^{(2)} = \frac{1}{2} \sum_{\bvec{k} \in \mathrm{BZ}} \psi_\bvec{k}^\dagger H(\bvec{k}) \psi_\bvec{k} + \mathrm{const.},
\end{equation}
where the bosonic spinor ${\psi}_\bvec{k} = (b'^x_{A,\bvec{k}},b'^x_{B,\bvec{k}},b'^x_{C,\bvec{k}},b'^y_{A,\bvec{k}}, \dots, (b'^x_{A,-\bvec{k}})^\dagger, \dots)^\top$ and ${H}(\bvec{k})$ is a $12 \times 12$ matrix.
The quadratic boson Hamiltonian \eqref{eq:H2biq} can be brought into normal form by means of a bosonic Bogoliubov transformation $\psi_\bvec{k} = T(\bvec{k}) \gamma_\bvec{k}$, where ${\gamma}_\bvec{k}$ is a spinor of normal modes. Given the size of $H(\bm{k})$, an analytical construction of $T(\bvec{k})$ is not feasible and we resort to the numerical algorithm by Colpa \cite{colpa}\footnote{See also Appendix A in Ref.~\cite{smit20} for an extended discussion.}.

\subsubsection{Fluctuation-induced moment reduction}
\label{sec:moment-reduc}

Fluctuations lead to a reduction of magnetic and quadrupolar moments compared to their classical (mean-field) values. 
These can be taken into account in a systematic manner in the flavor wave theory by computing observables to the first subleading order in $1/M$, analogous to $1/S$ corrections in the standard spin-wave theory for dipolar spins.

As the phases to be considered in general will have a mixed dipolar and quadrupolar character (i.e.\ the dipolar moment will be reduced $|\langle \vec S \rangle |^2 < 1$ already at the classical level), a systematic way to quantify the fluctuation-induced reduction of moments is to compute the renormalization of the $\uvec d$-vector to the first subleading order in $1/M$.

To this end, we first start with the spinor $\uvec{b} = (b_x,b_y,b_z)^\top$ and then condense an arbitrary linear combination of the three bosons, which leads to a $1/M$ expansion of the spinor expectation value $\langle \uvec{b} \rangle = \sqrt{M} \uvec{d}^{(0)} + M^{-1/2} \uvec{d}^{(1)}+ \mathcal{O}(1/M)$, where $\uvec{d}^{(0)}$ is the classical (mean-field) director with $|\uvec{d}^{(0)}|^2 = 1$.
While $ \uvec{b}^\dagger \uvec{b} = M$ holds as an operator identity to all orders, we can consider renormalizations of the classical amplitude
\begin{equation}
	\langle \uvec b^\dagger \rangle \langle \uvec b \rangle = M +  {\uvec d^{(0)}}^\dagger \uvec d^{(1)} + {\uvec d^{(1)}}^\dagger \uvec d^{(0)} + \mathcal{O}(M^{-1/2}).
\end{equation}
Using a unitary transformation $\uvec b \to U^\dagger \uvec b'$ to the canonical basis (the $b^z$ boson is condensed), we can then compute
\begin{equation} \label{eq:norm-bspin-exp}
	\langle \uvec b^\dagger \rangle \langle \uvec b \rangle = \langle {b_z'}^\ast \rangle \langle b_z' \rangle = M- \langle {b_x'}^\dagger {b_x'} \rangle - \langle {b_y'}^\dagger b_y' \rangle,
\end{equation} 
where the operator bilinears can be evaluated using the Bogoliubov transformation that diagonalizes Eq.~\eqref{eq:H2biq}.

While it is tempting to generalize the above procedure and evaluate the expectation value $\langle \uvec b \rangle$ directly by using the unitary matrix  $\umat{U}$ to transform to the canonical basis (which would facilitate a straightforward systematic $1/M$ expansion of all observables, such as the total magnetization), we emphasize that in general, this requires the use of non-linear flavor-wave theory:
In general, the matrix $U$ depends on variables which parametrize the states, i.e.\ $\umat{U} = \umat{U}(\eta_1, \dots)$.
These may e.g.\ allow one to continually tune from a purely dipolar state to a purely quadrupolar state and can be thought of as generalized canting angles.
As in conventional spin-wave theory, these angles are to be expanded in a power series in $1/M$, $\eta = \eta^{(0)} + M^{-1} \eta^{(1)} + \dots$, and consequently the unitary $U = U^{(0)} + M^{-1} U^{(1)} + \dots$ receives $1/M$ corrections which can be determined by considering corrections to the harmonic ground state due to boson interactions contained in $\mathcal{H}^{(3)}$ \cite{coletta12,consoli20}.
Combining these results, the spinor expectation value reads to first subleading order 
\begin{align}
	\langle \uvec b\rangle &= \left(\umat{U}^{(0)} + M^{-1} \umat{U}^{(1)} + \dots\right)^\dagger \langle b'_z \rangle \nonumber\\ &= \sqrt{M} {\umat{U}^{(0)}}^\dagger \uvec d'^{(0)} + \frac{1}{\sqrt{M}} \left[ {\umat{U}^{(0)}}^\dagger \uvec d'^{(1)} + {\umat{U}^{(1)}}^\dagger \uvec{d}'^{(0)} \right] + \dots,
\end{align}
with the second term in the brackets resulting from corrections of the angular variables.
Of course, these corrections vanish in those phases where no odd-boson terms are allowed in the Hamiltonian by symmetry, such as ferromagnetic or ferroquadrupolar states.

Given the complex nature of some of the phases encountered, and in the interest of generality we refrain from undertaking the non-linear flavor-wave computation sketched above.
We again emphasize that the preceding discussion applies to \emph{angular} corrections to $\uvec d$ (where the angles now also might tune between quadrupolar and dipolar spin states), and that the norm $|\langle \uvec b \rangle|^2 = \langle \uvec b^\dagger \rangle \langle \uvec b \rangle$ uniquely quantifies the renormalization of the amplitude of the $\uvec d$-vector due to quantum fluctuations, which can also be understood by noting the identity
\begin{align}
	\frac{1}{2} \langle Q^{\alpha \beta} \rangle \langle Q^{\alpha \beta} \rangle + \langle S^\alpha \rangle \langle S^\alpha \rangle = \frac{2}{3}-\frac{4}{3} |\langle \uvec{b} \rangle|^2 + 2 |\langle \uvec{b}\rangle|^4 
	\nonumber\\ \equiv \frac{4}{3} \quad \text{for }M=1\text{ and }|\langle \uvec{b} \rangle| = 1,
\end{align}
which fixes the ``total amplitude'' of joint spin and quadrupolar components, and will become renormalized for $|\langle \uvec b \rangle|^2 < 1$.

\section{Phase diagrams}
\label{sec:disc-phase-diagr}

\begin{figure*}[htbp]
	\includegraphics[width=.9\textwidth]{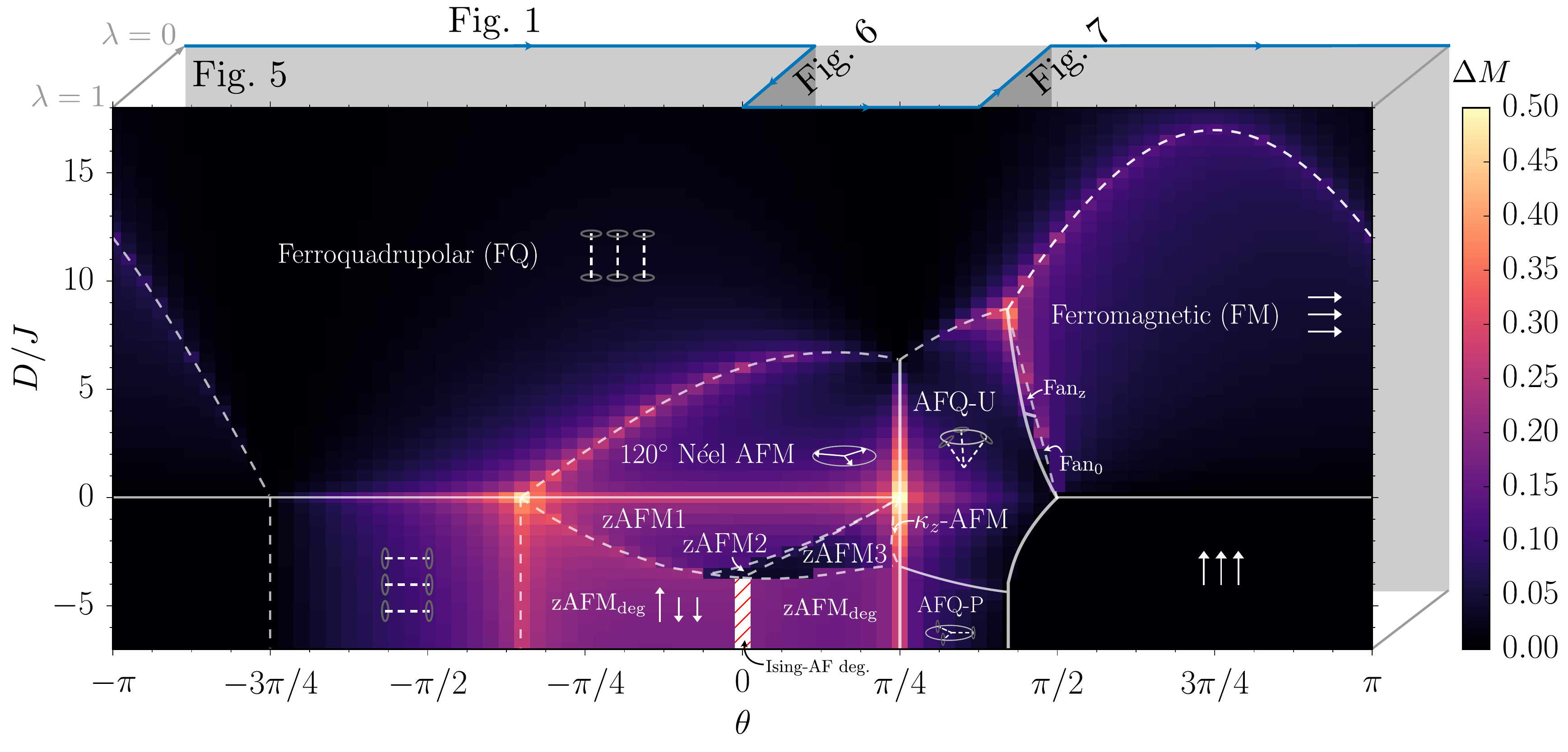}	
	\caption{Color plot of the quantum fluctuations in the classical ground state phase diagram obtained by minimizing the Hamiltonian of the bilinear-biquadratic model for couplings $J_1 = J \cos \theta$ and $J_2 = J \sin \theta$ and single-ion anisotropy $D$ ($D>0$ (resp.\ $D<0$) represents easy-axis (resp.\ easy-plane) anisotropy). The third direction represents the XXZ anisotropy in the bilinear and biquadratic terms ($\lambda=1$ (resp.\ $\lambda=0$) is the isotropic (resp.\ Ising) limit). Solid (dashed) lines indicate first-order (second-order) phase transitions. A description (and exemplary parametrization in terms of directors) of the various states encountered is given in Tab.~\ref{tab:states}. The blue line depicts the path in the $\theta$-$\lambda$-plane shown in Fig.~\ref{fig:alltogether}.}
	\label{fig:pd_theta_D}
\end{figure*}

\begin{figure}
	\centering
	\includegraphics[width=.8\columnwidth]{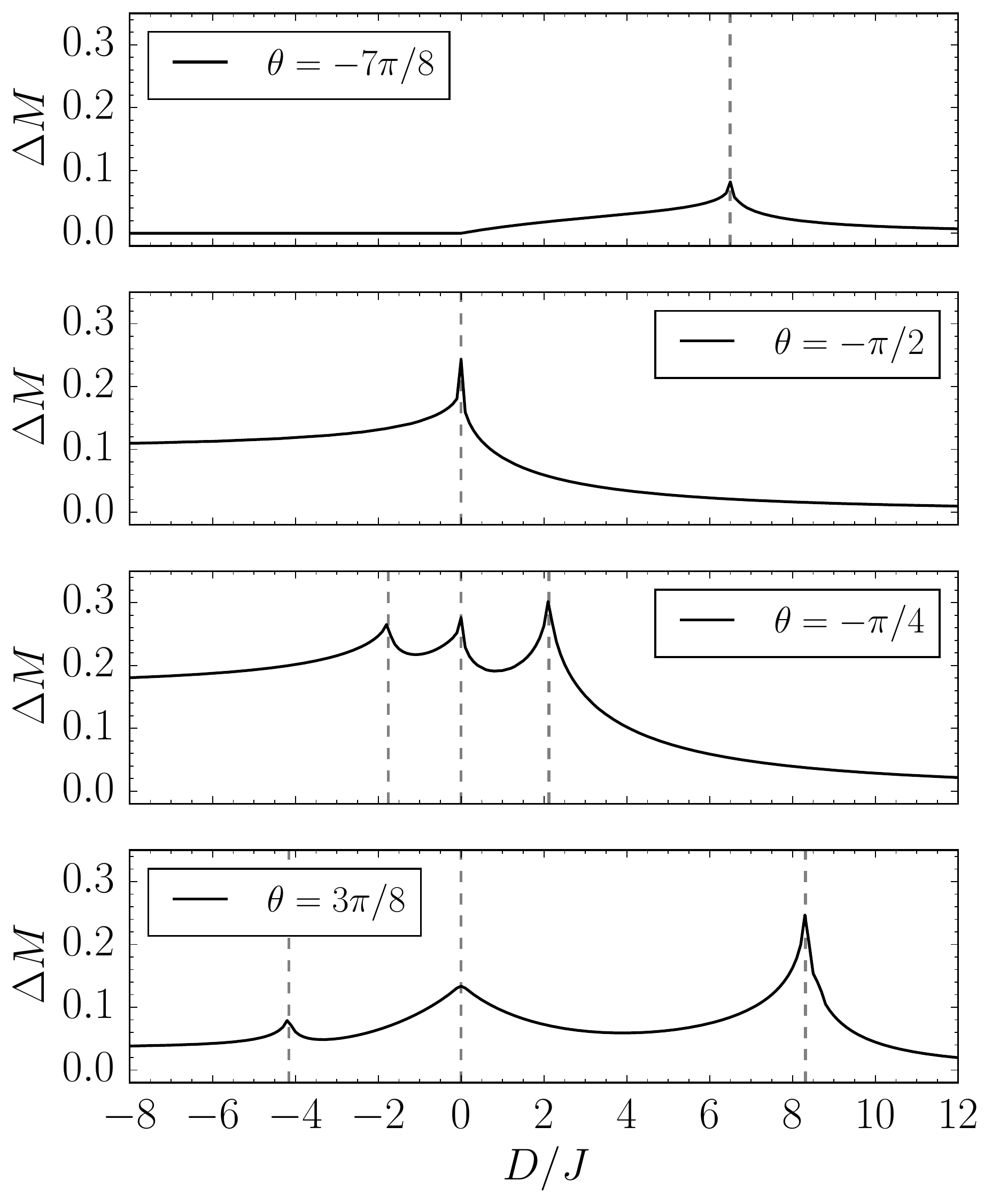}
	\caption{\label{fig:cutsAtLambda1} Correction $\Delta M$ to the classical order parameter amplitude obtained in linear flavor-wave theory, as a function of the anisotropy $D/J$ for isotropic exchange interactions $\lambda = 1$ and selected values of $\theta$. Grey dashed lines indicate phase boundaries.}
\end{figure}

When constructing phase diagrams, we assume that all classical ordering patterns have a three-sublattice structure, so that finding the classical ground state amounts to minimizing the $\sim M^2$ contribution of the Hamiltonian, which, using the parametrization \eqref{eq:dvec-angles} for the $\uvec d$-vector on the three sublattices, is a function of $3\times 4$ angular variables.
In practice, we find it easier to perform symbolic manipulations to expand the Hamiltonian in powers of $1/M$ for a general $\uvec{d}$-vector (which enters through the unitary basis change to the canonical basis) and then numerically minimize the classical piece $\mathcal{H}^{(0)}$ using \texttt{Mathematica}.
In order to construct the phase boundaries, we have found minimal parametrizations (i.e.\ $\uvec{d}$-vectors with as few parameters as possible to represent at least one state belonging to the degenerate manifold of each symmetry-broken phase, see also Tab.~\ref{tab:states}) and optimise these parameters for a given set of $D,\theta$ and $\lambda$.
This allows us to subsequently construct phase boundaries (which, in simple cases, can be determined analytically, but in general require the numerical solution of an implicit equation). 

\subsection{Isotropic exchange with single-ion anisotropy: Comparison with Moreno \textit{et al.}, Fig.~\ref{fig:pd_theta_D}}
\label{sec:iso-moreno}

The resulting phase diagram (i.e.\ at $\lambda=1$) as a function of $\theta$ and $D$ is shown in Fig.~\ref{fig:pd_theta_D}, matching previous results by T\'oth \cite{tothesis} as also discussed by Moreno \textit{et al.} in Ref.~\onlinecite{moreno14}.
We list all states found, with an illustration of the configuration of the dipolar moments and/or directors and a description in Tab.~\ref{tab:states}, as well as examples for ``minimal'' representatives.

We briefly discuss the occurring phases and their defining characteristics:
In general, we find that for sufficiently large positive $D > 0$ the system is in a ferroquadrupolar phase with the directors aligned along the anisotropy axis, $\uvec d \parallel \hat{z}$, consistent with the fact that for $D \gg |J_1|,|J_2|$ the product wavefunction $\prod_i \ket{0}_i$ is an eigenstate of the Hamiltonian $\mathcal{H}$.
Conversely, for $D < 0$ and $|D| \gg |J_1|,|J_2|$, the single-ion anisotropy does not select a unique ground state, but only a submanifold of product states with easy-axis character, $\prod_i (\alpha_i \ket{\uparrow}_i + \beta_i \ket{\downarrow}_i)$. The nature of the ordering for large $-D > 0$ thus necessarily depends on the nature of the exchange couplings $J_1, J_2$ (and $\lambda$) which select certain states of the above submanifold.

For purely ferromagnetic exchange at $\theta = \pi$, the system orders ferromagnetically (FM) with the spins lying in the $\hat{x}\hat{y}$-plane or parallel to the $\hat{z}$-axis for $D> 0$ or $D<0$, respectively.
Increasing $\theta$, the strength of ferromagnetic interactions become smaller, so that for $\theta \geq - 3 \pi/4$ we find ferroquadrupolar ordering (which breaks the $\SO(3)$ spin rotation symmetry) even at $D=0$.
Beyond $\theta = - \pi/2$, dipolar exchange interactions become stronger and are antiferromagnetic $J_1 < 0$, inducing a plethora of antiferromagnetically ordered states for $-\arctan(2) \leq \theta \leq \pi/4$:
For $D > 0$, a $120^\circ$ N\'eel-ordered state is found with dipolar moments lying in the $\hat{x}\hat{y}$-plane.
The spin length $|\langle \vec S \rangle |^2$ is reduced upon approaching the phase boundary to FQ at large $D> 0$.
On the other hand, for sufficiently small $-D > 0$, several coplanar antiferromagnetic states are found: While zAFM1, stabilized primarily for ferroquadrupolar exchange at $\theta <0$ exhibits a ``Y''-like configuration with one moment aligned along the easy-axis, this moment is rotated away from the $\hat{z}$-axis for zAFM2 which emerges for $\theta >0$. zAFM3 also shows a ``Y''-like configuration, with one moment lying in the $\hat{x}\hat{y}$-plane and the other two legs rotated out of the plane in a symmetric manner.
In addition, there is a small region in the phase diagram (``$\kappa_z$-AFM'') which features a coplanar spin configuration with the plane common to all three spins rotated away from the $\hat{x}\hat{y}$-plane, yielding a finite $\hat{z}$-component $\kappa_z$ of the spin chirality $\vec{\kappa} \sim \vec S_A \times \vec S_B + \vec S_B \times \vec S_C + \vec S_C \times \vec S_A$.

Upon increasing $-D>0$, mean-field theory yields a manifold of states with dipolar order along the $\hat{z}$-axis, as it is energetically favorable for mean-field states to have vanishing $d^z = 0$ on each site, tantamount to restricting the Hamiltonian to $\mathrm{Span}(\ket{+1},\ket{-1})$. Projected to this subspace, the Hamiltonian takes the form, as pointed out by Refs.~\onlinecite{tothesis,moreno14},
\begin{align}
  \label{eq:1}
  &\mathcal{P}_{S^z= \pm1} {\mathcal{H}|}_{\lambda=1} \mathcal{P}_{S^z= \pm1}\\
  &=\sum_{\langle i j \rangle} \big[ \frac{J_2}{2} \left(\sigma^x_i \sigma^x_j + \sigma^y_i \sigma^y_j \right) + \left(J_1 - \frac{J_2}{2} \right) \sigma^z_i \sigma^z_j \big] + \mathrm{const},\nonumber
\end{align}
i.e.\ that of an effective triangular lattice XXZ model (for $\theta\neq 0,\pi$) for the $\ket{\uparrow,\downarrow} \equiv \ket{\pm 1}$ pseudospin-1/2 degree of freedom. At $\theta=0$, i.e.\ for zero quadrupolar interactions and antiferromagnetic dipolar ones (recall $J_1=J\cos\theta$, $J_2=J\sin\theta$), $\mathcal{H}_{|\lambda=1}$ projects to the AFM Ising model which, as discussed above, is well-known to host a macroscopically degenerate set of states, the triangular Ising AFM states, a representative of which is shown in Fig.~\ref{fig:side}(b). We find, like others,\cite{henley92,murthy97,tothesis,moreno14} that at the {\em classical} level, the ground state of this model both for $-\arctan2<\theta<0$ and $0<\theta<\pi/4$ exhibits an \emph{accidental} global $\Uone$ degeneracy not related to physical symmetry operations (i.e.\ it is not the U(1) symmetry of the XXZ model). Murthy et al.~\cite{murthy97} showed that this accidental degeneracy could be lifted by a quantum order-by-disorder mechanism. Quantum mechanically, this model was found to host supersolid phases. We further comment on both aspects, as well as on the $\theta=0$ model, in Sec.~\ref{sec:region-A}. 

Increasing $\theta$ beyond $\theta = \pi/4$, interactions become dominantly antiferroquadrupolar ($J_2 >0$) and, for sufficiently large $-D > 0$, stabilize a planar antiferroquadrupolar configuration (AFQ-P) for which the directors lie in the $\hat{x}\hat{y}$-plane and form relative angles of $120^\circ$ with each other %
\footnote{Note that for purely quadrupolar configurations, one can employ the $\Uone$ phase degree of freedom in defining the directors to choose purely real $\vec d$. This further implies that the global sign of the $\vec d$-vectors is undetermined, and the respective angles of directors are only well-defined modulo $180^\circ$.}.
On the other hand, for small $-D> 0$ as well as small $D > 0$, we find an ``umbrella'' configuration of the $\vec{d}$-vectors which closes up as one progresses from $-D >0$ to large $D>0$, eventually entering the FQ phase via a second-order phase transition. Notably, at the $\SO(3)$-symmetric point $D=0$ (recall $\lambda=1$ here), the directors are orthogonal to each other \cite{laeuch06}.

Further increasing $\theta$, these antiferroquadrupolar phases give way to the ferromagnetic state described earlier.
While there is a direct AFQ-FM transition for $D<0$, T\'oth has uncovered a series of fan-like spin configurations (``Fan$_{0,z}$'') for $D>0$ with a spontaneous magnetisation in the $\hat{x}\hat{y}$-plane.

\begin{figure}
	\centering
	\includegraphics[width=.9\columnwidth]{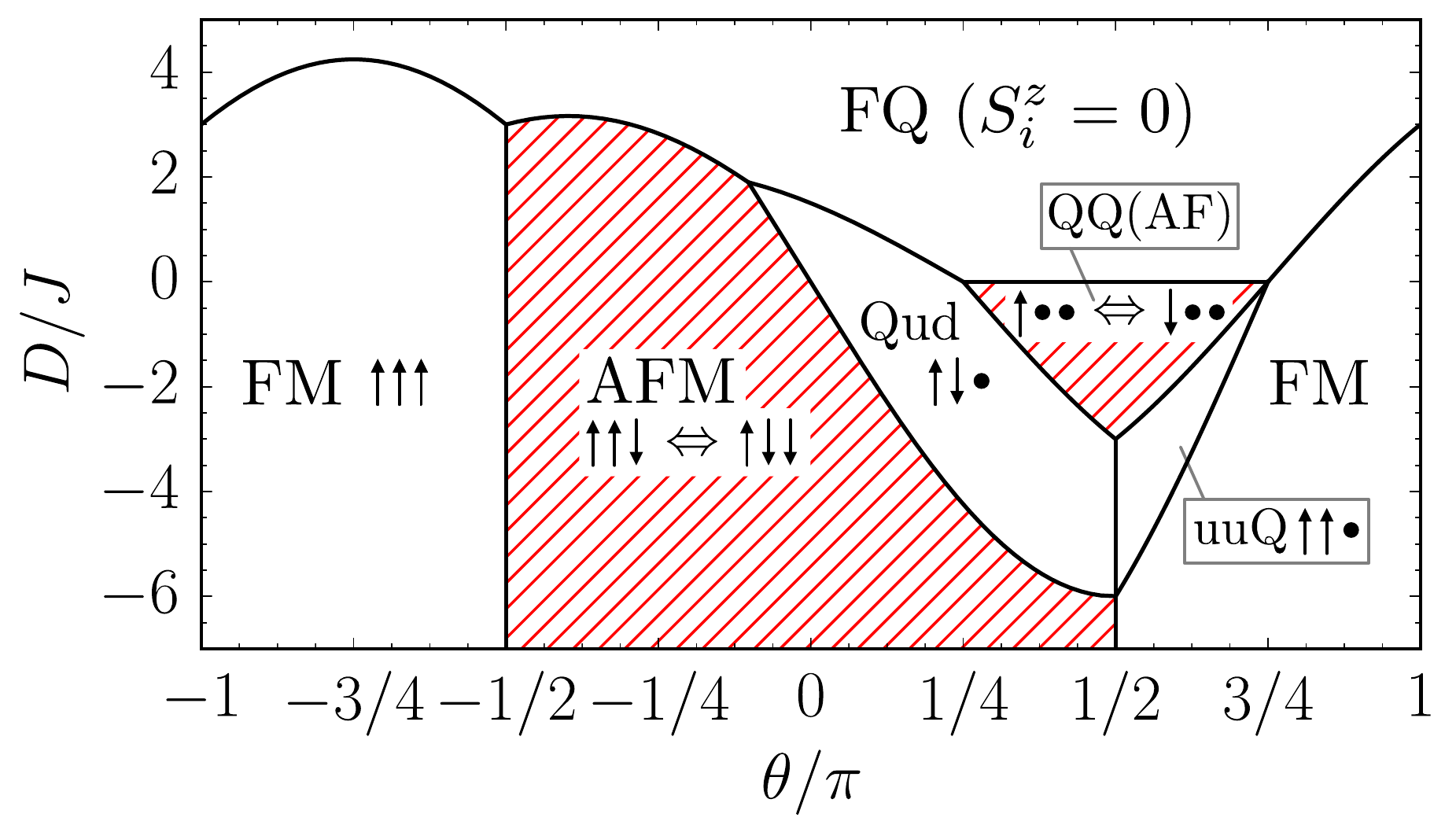}
	\caption{\label{fig:pd_ising}Ising limit ($\lambda=0$) phase diagram of $\mathcal{H}|_{\lambda=0}$ as a function of $\theta$ and $J$. The symbols indicate the orientation of three spins $S^z$ belong to a triangular plaquette, with the up-/down-arrows denoting $\ket{S^z=\pm 1}$, and a black dot denoting $\ket{S^z=0}$. Note that the ferromagnetic (FM, $\uparrow\uparrow\uparrow$) and 2/3 ferromagnetic ($\uparrow\uparrow\!\bullet$) states are twofold degenerate by $\mathbb{Z}_2$ symmetry. The red hatching indicates phases with a macroscopic ground-state degeneracy. All phase transitions shown are purely due to level crossing and of first order.} 
\end{figure}

\subsection{Longitudinal interactions: Exact ground states in the Ising (Blume-Capel) limit, Fig.~\ref{fig:pd_ising}} \label{sec:exact-Ising-lim}

We now study the $\lambda =0$-limit of $\mathcal{H}$ in \eqref{eq:h0} for which transverse exchange interactions are absent and $S^z_i$ are good quantum numbers, admitting an exact solution.
For $\theta = 0,\pi$ (i.e.\ dipolar interactions), the model is usually referred to as the Blume-Capel model \cite{blume66,capel66} and has been studied on the triangular lattice using Monte Carlo simulations \cite{tribc13}.
In general, a finite $J_2 = J \sin \theta$ will either energetically favor or punish dipolar states (with $S^z = \pm 1$) on adjacent sites.
To map out the phase diagram as a function of $D/J$ and $\theta$ we note that the Hamiltonian can be rewritten in terms of a sum over triangular plaquettes,
\begin{align}
	\mathcal{H}|_{\lambda =0} = \sum_{\triangle}& \bigg[ \frac{J \cos \theta}{4} \left(S^z_\triangle\right)^2 + \frac{J \sin \theta}{4} \left((S^z)_\triangle^2\right)^2 \nonumber\\
	&- \left(\frac{J \left(\cos \theta + \sin \theta\right)}{4} - \frac{D}{6}\right)(S^z)_\triangle^2\bigg],
\end{align}
where the sum extends over \emph{all} triangles of the lattice (both up- and down-pointing).
Here, $S^z_\Delta = \sum_{j \in \Delta} S_j$ denotes the spin per triangle, and we define $(S^z)^2_\Delta := \sum_{j \in \Delta} (S^z_j)^2$.
By comparing energies of various spin configurations on each triangle, one can then map out the phase diagram shown in Fig.~\ref{fig:pd_ising}:
For any $\theta$, a sufficiently large positive single-ion anisotropy $D$ will eventually stabilize a ferroquadrupolar phase with $\ket{S^z=0}$ on every site.
As expected, we find a ferromagnetic phase around $\theta = \pi$.
At $\theta = -\pi/2$, the dipolar Ising spin interactions change signs and become antiferromagnetic, resulting in the well-known macroscopic degeneracy of $2^{N/3}$ (for a system of $N$ sites) spanned by states with two parallel and one antiparallel spins per triangle.
On the other hand, for $\theta > 0$ the antiferroquadrupolar $J_2 > 0$ disfavors configuration with adjacent dipolar spins which competes with the easy-axis anisotropy for $D<0$, resulting in a complex phase diagram, with novel states contingent on the $S=1$ nature of local moments, featuring:
\begin{enumerate}
	\item A mixed antiferromagnetic-quadrupolar state (``Qud'') which has $S^z \in (+1,-1,0)$ on the three sublattices of a $\sqrt{3} \times \sqrt{3}$ unit cell,
	\item partial ferromagnetism (``uuQ'') with $m^z = 2/3$ magnetization due to one of the three sublattices being in a $S^z=0$ state, and
	\item a macroscopically degenerate phase (``QQ(AF)'') with dipolar moments $S^z=0$ vanishing on two sublattices, and the third sublattice satisfying the easy-axis constraint $(S^z)^2 =1$, admitting any linear combination $\alpha \ket{+1} + \beta \ket{-1}$ with $\alpha,\beta \in \mathbb{C}$ as a local wavefunction on that site, giving rise to a (quantum) pseudospin-1/2 degree of freedom per triangular plaquette. Note that at this stage, these resulting pseudospins-1/2 are {\em not} required to be aligned along the $\hat{z}$-axis, in contrast to the previously discussed Ising-like degeneracy in the dominantly antiferromagnetic phase. 
\end{enumerate}
As will be discussed below, some of the (mean-field) phases found for finite transverse exchange $\lambda \neq 0$ can be related to their $\lambda \to 0$ counterparts.

\subsection{Anisotropic transverse exchange}

Having discussed the mean-field phase diagrams for isotropic ($\lambda =1$) exchange interactions and the limit of longitudinal $S^z$-conserving interactions at $\lambda = 0$, we now turn to XXZ-type exchange interactions and briefly discuss occurring phases, with a more detailed discussion of phases with additional degeneracies relegated to Sec.~\ref{sec:degen}. The phase diagram as a function of $D$ and $\lambda $ in $J_1^{xy} = \lambda J_1^z$ where we fix $J_1^z \equiv J^z$ is shown for $\theta = 0$ in Fig.~\ref{fig:lambdaCut_0} and for $\theta= 3 \pi / 8$ in Fig.~\ref{fig:lambdaCut_3pi8}.
We stress that since for $\lambda = 0$ spin quantum numbers $S^z_i$ are conserved, the phase diagrams are exact in this limit ($\lambda=0$ corresponds to the left hand side axes in Figs.~\ref{fig:lambdaCut_0} and \ref{fig:lambdaCut_3pi8}).

\begin{figure}
	\includegraphics[width=.9\columnwidth]{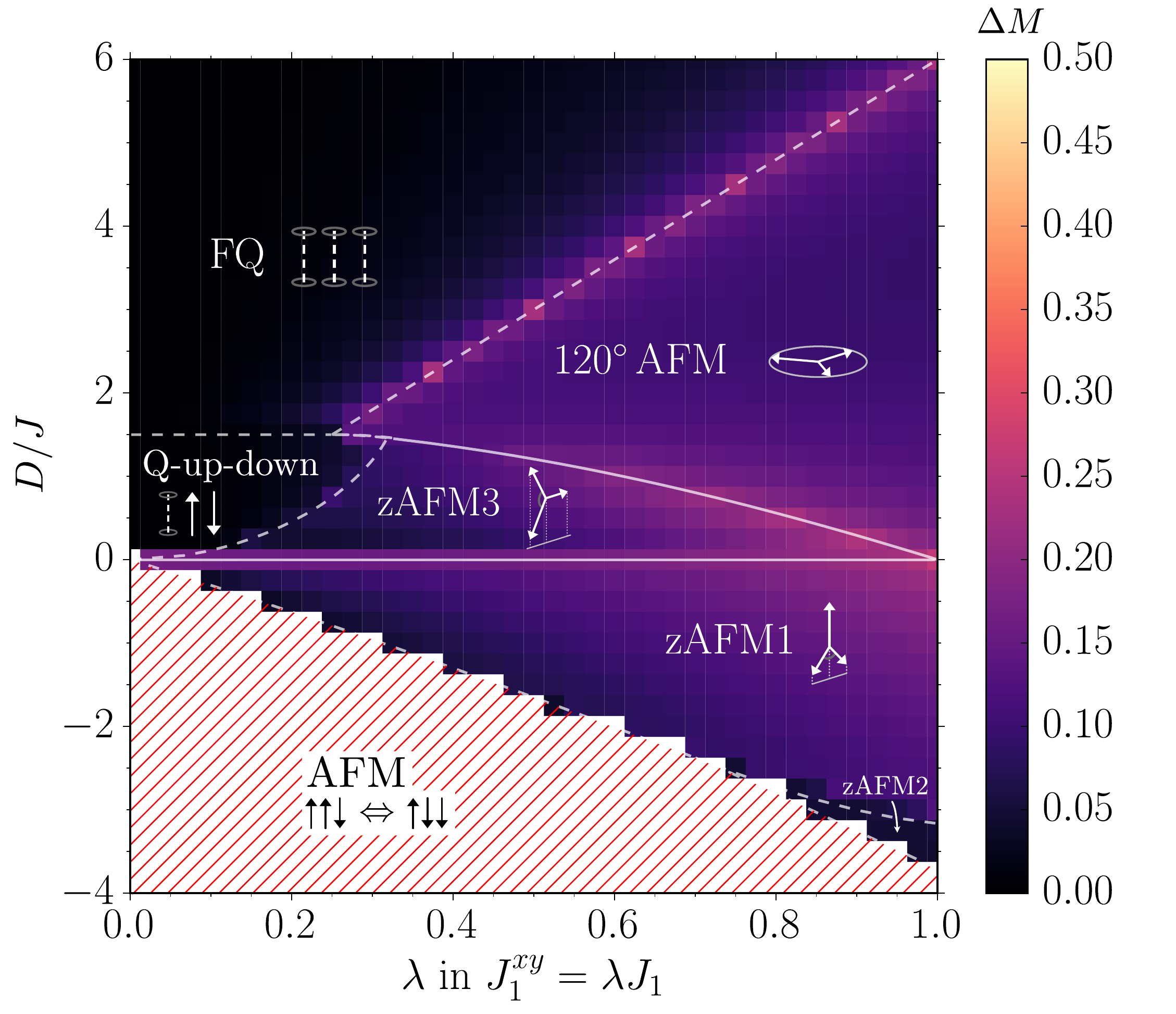}
	\caption{\label{fig:lambdaCut_0}$\theta = 0$ (bilinear) phase diagram obtained by variationally minimizing the mean-field Hamiltonian of the bilinear-biquadratic model for couplings $J_1 = J \cos \theta$ and $J_2 = J \sin \theta$ as a function of single-ion anisotropy $D$ and exchange anisotropy $\lambda$. Solid (dashed) lines indicate first-order (second-order) phase transitions.}
\end{figure}

\subsubsection{Dipolar antiferromagnetic exchange, $\theta = 0$, Fig.~\ref{fig:lambdaCut_0}}

Using an analogous line of thought as above, we consider the projection to the $S^z= \pm1$ manifold,
\begin{equation}
  \label{eq:2}
  \mathcal{P}_{S^z= \pm1} \mathcal{H}|_{\theta=0} \mathcal{P}_{S^z= \pm1} =\sum_{\langle ij\rangle}J\sigma_i^z\sigma_j^z,
\end{equation}
from which it is clear that the mean-field phase diagram hosts an antiferromagnetic Ising phase with a macroscopic ground-state degeneracy for sufficiently large $-D> 0$.
While this degeneracy is \emph{exact} in the $\lambda=0$-limit, it is an artefact of mean-field theory for any $\lambda > 0$.
Two mechanisms for lifting this degeneracy are discussed in Sec.~\ref{sec:region-A}.

For small $\lambda$, we find that at small $D> 0$ a three-sublattice (generalized) Ising state (``Qud'') is stabilized with two antiparallel moments pointing parallel to $\hat{z}$, and the third moment being in the quadrupolar state $\ket{z} \equiv \ket{S^z = 0}$.
Further increasing $D$, we find FQ as usual.
On the other hand, starting in the three-sublattice Ising state and increasing $\lambda$, we observe a second-order phase-transition to the easy-plane antiferromagnet with one fully polarized moment pointing along $\hat{z}$.
For the easy-plane anisotropies $D> 0$, the coplanar zAFM3 state is realized, with one moment lying in the $\hat{x}\hat{y}$-plane and the remaining two moments tilted out of the plane in a symmetric manner.
Further increasing $D$, there is a first-order transition to the in-plane $120^\circ$ AF phase (which is dominant for $D>0$ and $\lambda \to 1$), with the critical $D/J$ decreasing down to zero as one approaches $\lambda = 1$.
 
\begin{figure}[tb]
	\includegraphics[width=.9\columnwidth]{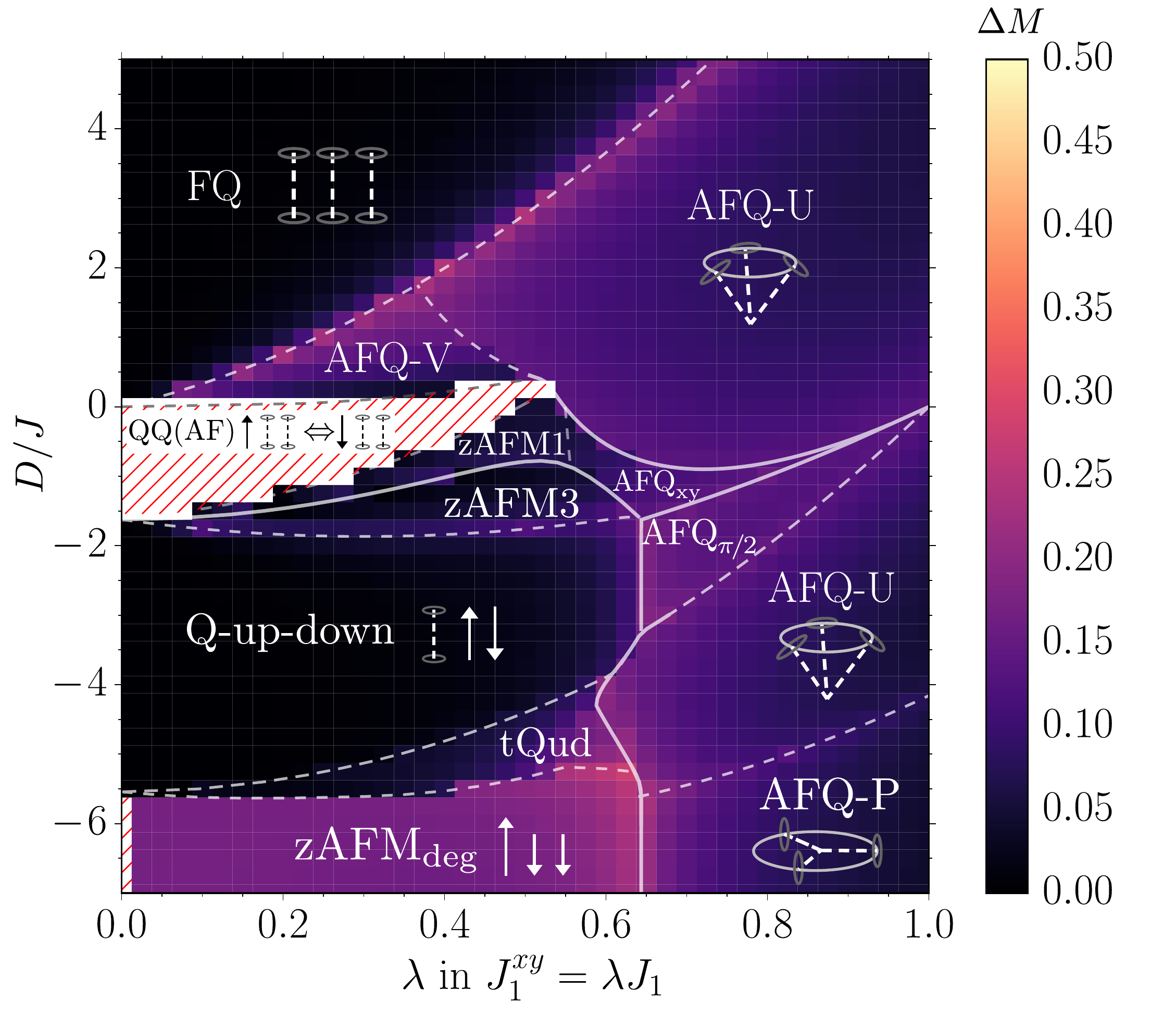}
	\caption{\label{fig:lambdaCut_3pi8}$\theta = 3 \pi / 8$ phase diagram obtained by variationally minimizing the mean-field Hamiltonian of the bilinear-biquadratic model for couplings $J_1 = J \cos \theta$ and $J_2 = J \sin \theta$ as a function of single-ion anisotropy $D$ and exchange anisotropy $\lambda$. Solid (dashed) lines indicate first-order (second-order) phase transitions.}
\end{figure}

\subsubsection{Mixed antiferromagnetic and quadrupolar exchange, $\theta = 3 \pi / 8$, Fig.~\ref{fig:lambdaCut_3pi8}}
\label{sec:mixed-anti}

For $\theta = 3 \pi/8$, interactions are dominantly antiferroquadrupolar $J_2 \approx 0.924 J$ with a weaker antiferromagnetic coupling $J_1 \approx 0.38 J$.
At $\lambda=1$, we have the aforementioned planar AFQ-P for strong easy-axis anisotropy, with a transition to an umbrella state, with the angle of the directors with the $\hat{z}$-axis closing as one approaches the phase boundary to the ferroquadrupolar phase (e.g. by increasing $D/J>0$).

Switching on the exchange anisotropy ($\lambda < 1$) the antiferromagnetic structure with mutually orthogonal directors becomes a stable phase AFQ$_{\pi/2}$ (as opposed to a singular point at $\lambda =1, D=0$), and an additional AFQ$_\text{xy}$ phase emerges which features one director laying in the $\hat{x}\hat{y}$-plane, and the remaining two vectors in an orthogonal plane (mirror symmetric about the $\hat{x}\hat{y}$-plane).
For strong exchange anisotropies $\lambda \lesssim 0.65$, we again encounter a phase, which we call zAFM$_{\rm deg}$, with an accidental ground-state degeneracy at strong $|D|$, and which can be seen to arise from the effective $S=1/2$ classical XXZ model obtained by projecting the Hamiltonian onto the $S^{z}=\pm1$ manifold, appropriate at infinite $|D|$, namely
\begin{align}
  &\mathcal{P}_{S^z= \pm1} \mathcal{H} \mathcal{P}_{S^z= \pm1} = \label{eq:proj-h}\\
  &\sum_{\langle i j \rangle} \big[  \frac{J_2 \lambda^2}{2} \left(\sigma^x_i \sigma^x_j + \sigma^y_i \sigma^y_j \right) + \left(J_1 - \frac{J_2 \lambda^2}{2} \right) \sigma^z_i \sigma^z_j \big] + \mathrm{const}.,\nonumber 
\end{align}
On a mean-field level, the $\lambda \to 0$ limit of this phase is singular and leads to macroscopically degenerate antiferromagnetic Ising ground states.
For intermediate easy-axis anisotropies, we find that the previously-mentioned three-sublattice Ising state with $S^z = +1,0,-1$, Q-up-down (Qud), is realized in a wide parameter regime.
This phase is separated from zAFM$_\text{deg}$ by a ``tilted'' (dubbed tQud) version of the three-sublattice Ising antiferromagnet. In this tQud the two $S^z = +1,-1$ moments are rotated away from the $\hat{z}$ axis by an identical angle (such that the system develops a spontaneous in-plane magnetization).
For intermediate $\lambda$ and small single-ion easy-axis anisotropies $-D> 0$, we further find that the easy-axis antiferromagnets zAFM1 and zAFM3 are stabilized.

Finally, remarkably, as one decreases $\lambda$ at small $D$, we encounter an Ising-like phase, QQ(AF), which features $S^z = 0$ on two sublattices and an easy-axis state $\alpha \ket{+1} + \beta \ket{-1}$ on the remaining third lattice, which gives rise to a pseudospin-1/2 degree of freedom.
Strikingly, a macroscopic ground-state degeneracy is found over an extended region in parameter space.
The latter extends down to the exactly solvable limit of $\lambda = 0$, which we can understand as follows.
Projecting the Hamiltonian to mean-field configurations of the form $S^z = (0,0,\pm1)$ on the three sublattices we find
\begin{equation} \label{eq:proj-trivial}
	\mathcal{P}_{0,0,\pm 1} \mathcal{H} \mathcal{P}_{0,0,\pm 1} \equiv \mathrm{const.}
\end{equation}
This implies that the extensive degeneracy found at $\lambda=0$ is not lifted in mean-field theory at finite $\lambda  >0$.
We can understand this in more physical terms by the fact that $\mathcal{H}$ contains only nearest-neighbor interactions while second-nearest neighbor ones are required (on a mean-field level) to induce correlations of the pseudospin-1/2 moments on the third sublattice.
We discuss the lifting of this degeneracy in Sec.~\ref{sec:degen-3pi8}.

\section{(Accidental) degeneracies and their breaking}
\label{sec:degen}

Here we analyze in more detail the phases for which our mean-field approach predicts a macroscopic degeneracy.

\subsection{Antiferromagnetic Ising regime} \label{sec:region-A}

In this section, we discuss those phases in the phase diagram which are macroscopically degenerate and where the states at each site are spanned by the $S_i^z=\pm1$ manifold. First we provide a semiclassical analysis away from the Ising and pure dipolar limits. Then we promote our states and model to quantum ones, and study the fate of these states within perturbation theory. 

\subsubsection{Semiclassical analysis $\lambda \neq 0$ and $\theta \neq 0$, AFM$_{\rm deg}$ states}

Here we address the states which we labeled zAFM$_{\rm deg}$ in the phase diagrams. As discussed in Secs.~\ref{sec:mixed-anti} and \ref{sec:iso-moreno}, they can be understood as emerging from the projection of $\mathcal{H}$ into the $S^{z}=\pm1$ subspace, which we give for any $\theta$ and $\lambda$ in Eq.~\eqref{eq:proj-h}. While at $\theta=0$ or $\lambda=0$, Eq.~\ref{eq:proj-h} is an effective $S=1/2$ Ising model with macroscopic degeneracy, for any $\lambda > 0$, antiferroquadrupolar exchange interactions present for any $\theta \neq 0$ lift this macroscopic degeneracy at the mean-field level, since Eq.~\eqref{eq:proj-h} contains transverse exchange terms.

In the parameter regimes where we have identified a zAFM$_{\rm deg}$ state, the {\em classical} (spin-$S$, with $S\gg1$) XXZ Hamiltonian Eq.~\eqref{eq:proj-h} on the triangular lattice features an accidental $\Uone$ degeneracy \cite{henley92,murthy97} of relative rotations of the dipolar pseudospins, which does not correspond to a physical symmetry operation of the model. This accidental degeneracy is broken by considering spin-wave corrections to the classical ground-state, as elucidated by Murthy, Arovas and Auerbach \cite{murthy97}.
In particular, these authors find that the (semi-)classical ground state selected by spin-wave corrections corresponds to a state which maximizes the total magnetization $m^z \sim \sigma^z_A + \sigma^z_B + \sigma^z_C$. They show that the extremization of the magnetization is a necessary condition for the linear spin-wave spectrum to contain \emph{two} zero-modes at momentum $\bvec k = 0$, corresponding to the spontaneously broken generators of \textit{(i)} the $\Uone$ in-plane spin rotation symmetry (i.e.\ a Goldstone mode) and \textit{(ii)} the accidental $\Uone$ degeneracy.
Conversely, for non-extremal $m^z$, the physical and accidental $\Uone$ degeneracies mix and corresponding zero-modes are no longer linearly independent, so that only one zero mode is found in the spectrum.
Murthy et al.\ argue that the former case is favorable to the latter, since each zero mode ``pins'' the dispersion to low energies, leading to a smaller spin-wave correction to the ground-state energy (recall $\Delta E^{(1)} \sim \sum_{\bvec{k},\mu} \omega_\mu(\bvec k)$).

An important corollary, as pointed out by T\'oth \cite{tothesis} and Moreno \textit{et al.}, of the above discussion of the effective dipolar mean-field XXZ model is that the stabilized semi-classical ground state exhibits \emph{both} in-plane $\Uone$-symmetry breaking order (of quadrupolar components) \emph{and} three-sublattice modulated longitudinal order of the out-of-plane (dipolar) $S^z$ spin components, thus exhibiting \emph{supersolidity}.

Finally, we note that, in order to select the order-by-disorder-favored state obtained by Murthy, Arovas and Auerbach as a {\em unique} ground state within the full $S=1$ variational mean-field ground-state search and subsequently compute the flavor-wave spectra and moment corrections shown in Fig.~\ref{fig:pd_theta_D} and in Sec.~\ref{sec:spectr}, we use the fact that the magnetization $m^z$ in the effective model can be extremized by applying an infinitesimal magnetic field. Indeed, since $\mathcal{P}_{S^z=\pm 1} S^z_i \mathcal{P}_{S^z=\pm 1} \equiv \sigma^z_i$, it is the ground state of $\lim_{h\to0} \mathcal{P}_{S^z=\pm 1} \left[ \mathcal{H}  - h\sum_i S^z_i\right] \mathcal{P}_{S^z=\pm 1}$. We indeed favor this approach over the inclusion of extra biquadratic interactions which can sometimes mimic order-by-disorder (``ObD'') ground-state selection mechanisms \cite{henley92}.

\subsubsection{Region $\mathcal{S}$: Perturbation theory in $J/D \ll 1$}
\label{sec:pert-JDll1}

Next, we formally consider the limit $D/J \to \infty$ and consider a quantum model.
In this limit, each site corresponds to a two-fold degenerate pseudospin-1/2 degree of freedom given by $\{\ket{S^z = +1},\ket{S^z=-1}\}$, and one can derive an effective Hamiltonian by performing perturbation theory in $J/D \ll 1$ \cite{tothesis}.
At first order, the effective Hamiltonian just corresponds to the projected Hamiltonian $\mathcal{H}_\mathrm{eff}^{(1)} = \mathcal{P}_{S^z=\pm 1} \mathcal{H} \mathcal{P}_{S^z=\pm 1}$ given in \eqref{eq:proj-h}. Notably, as mentioned before, this Hamiltonian features a macroscopic ground-state degeneracy in the limits $\theta = 0$ or $\lambda = 0$.
While the latter case matches the result of the exact analysis in Sec.~\ref{sec:exact-Ising-lim}, the former case is an artefact of truncating perturbation theory at first order in $J/D$.
At second order in $J/D$, there is a pseudospin exchange process on a given $\langle ij \rangle$-bond via intermediate $\ket{0_i, 0_j}$ states, giving rise to the effective Hamiltonian
\begin{align}
	\mathcal{H}^{(2)}_\mathrm{eff} = \frac{J^2 \lambda^2 (\cos \theta - \sin\theta)^2}{4 |D|} \sum_{\langle ij \rangle} \big[-\sigma^x_i \sigma^x_j &- \sigma^y_i \sigma^y_j + \sigma^z_i \sigma^z_j \big] \nonumber\\  &+ \mathrm{const.},
\end{align}
which is again of XXZ form (first given in Ref.~\onlinecite{tothesis}), so that, {\em up to} second order in $J/|D|\ll1$,
\begin{equation}
  \label{eq:3}
 \mathcal{H}^{(1)}_\mathrm{eff}+ \mathcal{H}^{(2)}_\mathrm{eff}=\sum_{\langle ij\rangle}\left[\hat{J}^{xy}(\sigma_i^x\sigma_j^x+\sigma_i^x\sigma_j^x)+\hat{J}^{zz}\sigma_i^z\sigma_j^z\right],
\end{equation}
with
\begin{eqnarray}
  \label{eq:4}
  \hat{J}^{xy}&=&J\frac{\lambda^2}{2}\left(\sin\theta-\frac{J}{|D|}\frac{1-\sin2\theta}{2}\right),\nonumber\\
  \hat{J}^{zz}&=&J\cos\theta-\hat{J}^{xy}.
\end{eqnarray}
In particular, the transverse exchange coupling is finite even for purely antiferromagnetic exchange $\theta = 0$.

The {\em quantum} $S=1/2$, XXZ model on the triangular lattice, which Eq.~\eqref{eq:3} describes, can be mapped to hardcore bosons with kinetic energy $t$ and repulsive density-density interactions $V$.
Previous studies have utilized a combination of analytical and numerical arguments \cite{melko05,wang09,jiang09} to provide evidence for the realization of an extensive phase with \emph{supersolid} order.
In particular, for frustrated hoppings $-1/2 \leq t/V \leq 0$ of the effective model, corresponding to $0 \leq \theta < \pi/4$ in our model, Wang \textit{et al.}\ \cite{wang09} find superfluid order parameters $\langle \sigma^+ \rangle \sim (m_\perp,-m_\perp,0)$ on the three sublattices and a weakly ferrimagnetic solid order $\langle \sigma^z \rangle \sim (-m^z,-m^z, 2 m^z- \delta)$ with $\delta > 0$, while for unfrustrated $0 \leq t/V \leq 0.1$ ($-0.15 \pi \leq \theta \leq 0$) the in-plane superfluid order is predicted to acquire a non-zero average, taking values in $(m_\perp, m_\perp, m_\perp)$ on the three sublattices.
Note that while the upper critical $\theta$ for the supersolid phase matches the classical found phase boundary $\theta = \pi/4$, the lower critical $\theta \approx 0.15 \pi$ suggested by Wang \emph{et al.}\ \cite{wang09} deviates from the classical phase boundary $\theta= -\arctan2 \approx -0.35 \pi$, possibly due to renormalization of coupling constants and induced longer-ranged couplings at higher order in perturbation theory. 

The applicability of these results for the first-order, pseudospin-1/2, XXZ model on the triangular lattice have been corroborated by Moreno-Cardoner \textit{et al.} using cluster mean-field methods \cite{moreno14}, in particular the distinct nature of supersolid phases for $\theta < 0$ and $\theta >0$, respectively. 
Interestingly, they also find that the lower critical $\theta$ is shifted compared to the classical value, while the upper critical value of $\theta = \pi/4$ appears to be stable.

\subsubsection{Region $\mathcal{A}$: Perturbation theory in $\lambda \ll 1$ at fixed $D,\theta$}
\label{sec:lambdall1}

\begin{figure}
	\centering
	\includegraphics[width=\columnwidth]{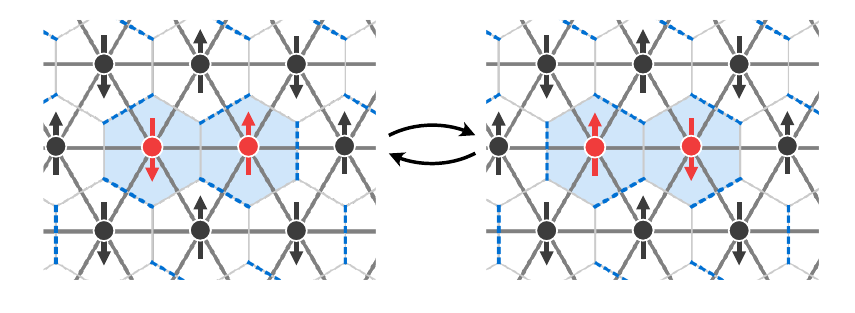}
	\caption{\label{fig:flippable-pair}Flippable pair\cite{fazand74} and two-plaquette resonance process relevant in perturbation theory in region $\mathcal{A}$ (finite $D<0$ and $\theta$, $\lambda\ll1$): Flipping two neighbouring $S^z=\pm 1 \to S^z=\mp 1$ yields off-diagonal matrix elements in the degenerate subspace of $|S_\triangle^z | = 1$, at order $\lambda^2$ in perturbation theory in $\lambda\ll1$. Also shown is the mapping to a dimer (dashed blue lines) model on the dual honeycomb lattice, with the spin-flip corresponding to a two-plaquette resonance process (shaded blue).}
\end{figure}

One virtue of the model \eqref{eq:h0} is that tuning the exchange anisotropy $\lambda$ allows us to access the supersolid region at \emph{finite} $-D/J>0$, using perturbation theory on top of the (exactly solvable) fully frustrated Ising case at $\lambda = 0$.

To this end, recall that for $-D/J > 0$ and $\lambda = 0$ and a sufficiently wide range of $- \pi /2 \lesssim \theta \lesssim \pi/2$ the model has an extensive number of Ising ground states where the local moments are of dipolar nature, and the spin on each plaquette $S^z_\Delta = \pm 1$, cf.\ Figs.~\ref{fig:pd_ising} and \ref{fig:side}(b).
We employ a similar analysis to Fazekas and Anderson in noting that there exist off-diagonal matrix elements for the effective Hamiltonian corresponding to flipping two antiparallel spins on a bond $\langle ij \rangle$ if they belong to a ``flippable pair'' as shown in Fig.~\ref{fig:flippable-pair} (in general, starting with an Ising ground state and applying a spin-flip can lead to plaquettes with $S^z_\Delta = \pm 3$, i.e. take the system out of the degenerate ground-state manifold).
However, in contrast to the $S=1/2$ case analysed by Fazekas and Anderson, in the $S=1$ case a spin-flip process needs to occur at second order in $\lambda$, since $(S^+_i S^-_j)^2 \ket{\downarrow_i \uparrow_j} = S^+_i S^-_j \ket{0_i 0_j} = \ket{\uparrow_i \downarrow_j}$.
Due to our choice of $\lambda$-scaling in \eqref{eq:h0}, this implies that the spin-flip can either occur through a second order process $\sim \lambda^2 J_1^2 / \Delta E$ (with $\Delta E \sim - D$ denoting the energy of an intermediate state), or at first order in $J_2 \lambda^2$ using biquadratic exchange, or second-order cross terms $\sim (J_1 \lambda) J_2 \lambda / \Delta E$, yielding the matrix element of the effective Hamiltonian
\begin{equation}
	\braket{\uparrow_i \downarrow_j|\mathcal{H}_\mathrm{eff}|\downarrow_i \uparrow_j} = -\frac{(J_1 \lambda)^2 - 2 J_1 J_2 \lambda^2 + J_2^2 \lambda^2}{\Delta E} + J_2 \lambda^2,
\end{equation}
where $\Delta E= J_1 - 11 J_2 - 2D$ is the energy difference of a degenerate Ising ground state to one with a flippable pair on bond $\langle ij \rangle$ replaced by $\ket{0_i 0_j}$. Note that the minus sign of the second term on the denominator results from evaluating $S^+_i S^-_j S^z_i S^z_j$ on a flippable bond in an Ising configuration.

In addition to these off-diagonal terms, there are \emph{diagonal} energy shifts resulting from projecting $J_2 \lambda^2 S^+_i S^-_j S^-_i S^+_j + \hc$ into the Ising manifold, as well as second order processes of the form $\sim - J_1^2 \lambda^2\braket{\uparrow_i \downarrow_j|S^+_i S^-_j|0_i0_j}\braket{0_i0_j|S^-_i S^+_j| \uparrow_i \downarrow_j}/\Delta E$, where $i,j$ are \emph{not} required to be flippable.
While the former contribution simply shifts the energy of an antiferromagnetic Ising bond, the latter contribution  depends (via $\Delta E$) on the Ising configuration of the 8 spins surrounding the pair $\langle i j \rangle$, making it difficult to rewrite it as an effective interaction acting within the degenerate ground-state sector.
However, we note that for the parameter regime considered here (i.e.~dominantly antiferromagnetic interactions), the intermediate energy $\Delta E$ for above expression is smallest for flippable pairs: For flippable configurations, the excited state contains defect plaquettes of ``$\uparrow\!00$'' and ``$\uparrow\downarrow\!0$'' types, while applying $S^+_i S^-_j + \hc$ on a non-flippable bond (for which some pairs of the 8 neighboring spins are aligned ferromagnetically) leads to some defect triangles with ``$\uparrow\uparrow\!0$'' (or ``$\downarrow\downarrow\!0$'') rather than ``$\uparrow\downarrow\!0$''.
Since for dominantly antiferromagnetic interactions one has $E_{\uparrow\uparrow0} > E_{\uparrow\downarrow0} > E_{\uparrow\uparrow\downarrow} > 0$, it follows that $\Delta E$ is smallest for flippable configurations.
We hence expect that diagonal processes in the degenerate ground-state manifold maximize the number of flippable plaquettes.
Note, however, that above arguments imply that the degeneracy of the flippable pair of spins within any given plaquette is not lifted by these diagonal terms, in contrast to off-diagonal terms which give rise to finite matrix elements for the flippable pair.
 
Further progress can be made by employing a dimer representation on the dual honeycomb lattice.
A dimer is placed on a honeycomb bond if the triangular bond $\langle ij \rangle$ bisected by it is frustrated, i.e.\ $S^z_i S^z_j = + 1$.
The degenerate Ising ground states thus satisfy the constraint that each honeycomb vertex is connected to exactly one dimer.

The off-diagonal matrix element at order $\lambda^2$ described above thus correspond to a two-hexagon resonance process as shown in Fig.~\ref{fig:flippable-pair}.
Having established that transverse exchange at order $\lambda^2$ maps to resonance processes within the dimer model for the degenerate Ising ground states, we can now straightforwardly apply the results by Wang \textit{et al.} in Ref.~\onlinecite{wang09} who use the aforementioned dimer mapping to study supersolid phases of the $S=1/2$ XXZ model on the triangular lattice (we expect that aforementioned diagonal contributions will further stabilise flippable plaquettes and will renormalize phase boundaries accordingly).

\subsection{Region $\mathcal{B}$: Degenerate dipolar moments on one sublattice} \label{sec:degen-3pi8}

\begin{figure}
	\centering
	\includegraphics[width=\columnwidth]{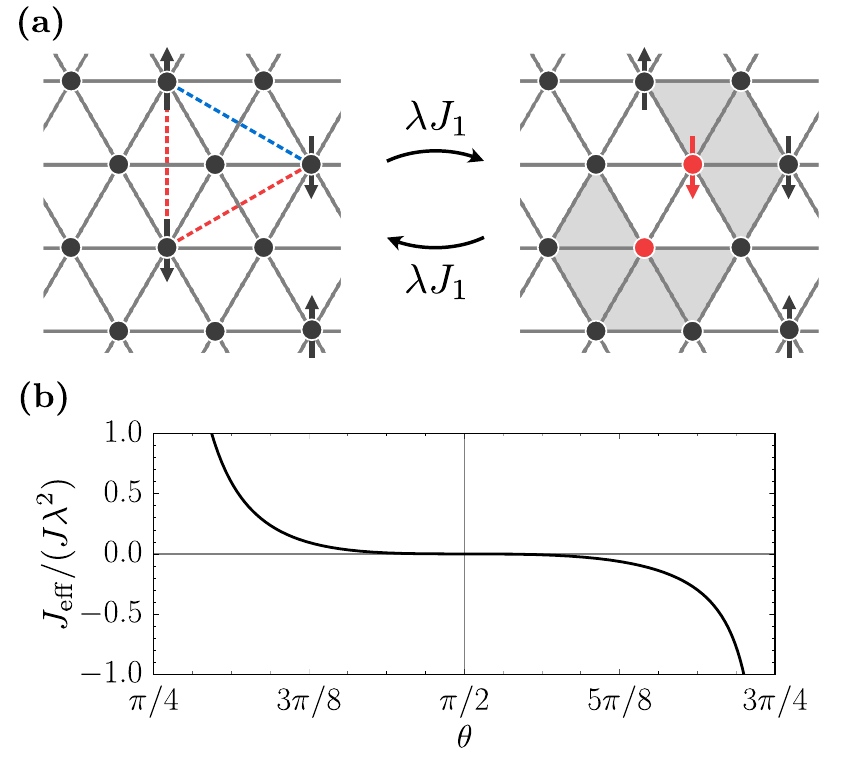}
	\caption{\label{fig:2nd-order-ising} Illustration of second-order processes in perturbation theory relevant in region $\mathcal{B}$ (QQ(AF) in the Ising limit, $\lambda\ll1$). (a) Second-order process giving rise to effective Ising interactions $J_\mathrm{eff}$ (dashed) among (second-neighbor) dipolar moments. Exchanging $S^z = \pm 1$ and $S^z=0$ on a bond gives rise to seven triangular plaquettes which are outside the ground-state manifold (shaded), and whose energy depends on the relative spin polarizations of the flipped spin and its neighbors. The process depicted gives identical matrix elements on the red dashed next-nearest-neighbor bonds, and further a matrix element between the unflipped neighbors (blue). These interactions are symmetrized when summing over all processes to obtain the full effective Hamiltonian. (b) Induced effective Ising coupling as a function of $\theta$. Note the sign change at $\theta = 0$.}
\end{figure}

For $\theta =3 \pi /8$ and sufficiently small $\lambda < 1$ and easy-axis anisotropies $-D>0$, variational mean-field theory predicts the presence of a macroscopically degenerate three-sublattice phase, with the moments on two sublattices being in a $S^z= 0$ state and the third sublattice featuring an arbitrary coherent superposition of $\ket{+1},\ket{-1}$, giving rise to pseudospin-1/2 degree of freedom.
As discussed, the degeneracy even at finite $\lambda$ found in variational mean-field theory is an artefact and follows from the triviality of the projected Hamiltonian \eqref{eq:proj-h} due to the nearest-neighbor nature of interactions in $\mathcal{H}$.

In turn, we use perturbation theory in $\lambda \ll 1$ to derive an effective Hamiltonian $\mathcal{H}_\mathrm{eff}.$ for the pseudospin-1/2 degree of freedom spanned by $\ket{\uparrow} = \ket{+1}$ and $\ket{\downarrow} = \ket{-1}$ on one of the sublattices, which is capable of lifting the aforementioned degeneracies.
For the second-order contribution, we consider a cluster of three pseudospins in a background of $S^z=0$ sites as shown in Fig.~\ref{fig:2nd-order-ising}, where for concreteness we fix the pseudospins-1/2 to live on the $C$ sublattice.
It is convenient to write the perturbing Hamiltonian as
\begin{align}
	&\mathcal{H}_\lambda = \sum_{\langle ij \rangle} \bigg[ \frac{\lambda J_1}{2} \left( S^+_i S^-_j + S^-_i S^+_j \right) + \frac{\lambda^2 J_2}{4} \left(S^+_i S^-_j + S^-_i S^+_j\right)^2 \nonumber\\
	&+ \frac{\lambda J_2}{2} \Big( \left(S^+_i S^-_j + S^-_i S^+_j\right) S^z_i S^z_j + S^z_i S^z_j  \left(S^+_i S^-_j + S^-_i S^+_j\right) \Big)\bigg]. \label{eq:lambda-pert}
\end{align}
As each neighboring site of a pseudospin-1/2 is in a $\ket{0}$-state, we find that the last term in \eqref{eq:lambda-pert} has vanishing contributions to matrix-elements between the degenerate ground-state and excited states, since either $S^\pm S^z \ket{0} = 0$ or $S^z S^\pm \ket{\pm 1} = S^z S^\mp \ket{\pm 1} = 0$ holds on $\langle ij \rangle$-bonds emanating a $C$-sublattice site.

Hence, at second order in $\lambda$, the only process that contributes to the effective Hamiltonian consists of a ``spin flip'' process with matrix element $\mathcal{H}_\lambda \ket{\pm 1_i, 0_j} = J_1 \lambda \ket{0_i, \pm 1_j}$ and the subsequent reverse spin flip to the original configuration (here $i \in A,B$ and $j \in C$ sublattices).
Crucially, the energy of the intermediate state depends on the pseudospin configuration of the $C$ sublattice sites which neighbor the $A$ or $B$ sublattice site $j$ in the above matrix element.
We thus find, for processes in the three-spin cluster as depicted in Fig.~\ref{fig:2nd-order-ising}(a), matrix elements of the effective Hamiltonian $h^\mathrm{eff}_{01}$ obtained by acting with the term in $\mathcal{H}_\lambda$ on the $\langle 0 1 \rangle$-bond, 
\begin{subequations}\begin{align}
	\braket{\uparrow_1\uparrow_2\uparrow_3 | h^\mathrm{eff}_{01} | \uparrow_1 \uparrow_2 \uparrow_3} &= -\frac{J_1 ^2 \lambda^2}{2 (J_1+ J_2)} + \mathcal{O}(\lambda^3) \\
	\braket{\uparrow_1\uparrow_2\downarrow_3 | h^\mathrm{eff}_{01} | \uparrow_1 \uparrow_2 \downarrow_3} &= -\frac{J_1^2 \lambda^2}{2 J_2} + \mathcal{O}(\lambda^3)\\
	\braket{\uparrow_1\downarrow_2\downarrow_3 | h^\mathrm{eff}_{01} | \uparrow_1 \downarrow_2 \downarrow_3} &= -\frac{J_1^2 \lambda^2}{2 (J_2-J_1)}+ \mathcal{O}(\lambda^3),
\end{align}\end{subequations}
and further matrix elements related by symmetry.
Note that all off-diagonal matrix elements vanish at second order:
The application of $\mathcal{H}_\lambda$ on two adjacent bonds can give rise to an ``exchange'' matrix element at second order, however this would leave a single $S^\pm$ acting on a pseudospin, taking the system out of the degenerate ground-state sector.
With the above given matrix elements, $h_{01}^\mathrm{eff}$ can be rewritten as a sum over pairwise Ising interactions among the pseudospins forming the three-spin cluster.
Summing over all bonds, we finally find the effective Hamiltonian for the pseudospins to be given by
\begin{equation}
	\mathcal{H}_\mathrm{eff}^{(2)} =\lambda^2 \underbrace{\frac{J_1^3(J_1+2 J_2)}{2 J_2(J_2^2-J_1^2)}}_{:= J_\mathrm{eff}} \sum_{\langle i j \rangle_C}  \sigma^z_i \sigma^z_j + \mathrm{const.},
\end{equation}
where $\langle i j \rangle_C$ denotes summation over nearest-neighbor bonds of the superlattice formed by $C$ sublattice sites.

\begin{figure}
	\centering
	\includegraphics[width=\columnwidth]{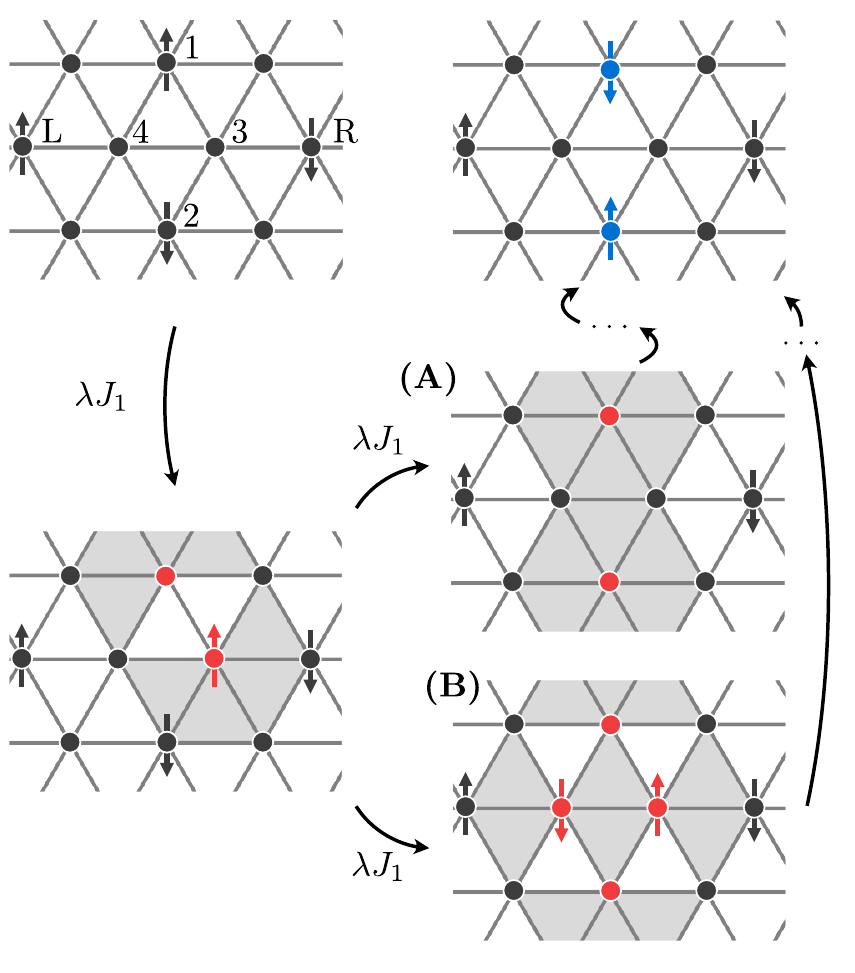}
	\caption{\label{fig:exge-4th-order} Example for fourth-order processes ($\lambda\ll1$) leading to transverse exchange couplings among pseudospins in region $\mathcal{B}$. While summing over all processes that include (A), i.e.\ $\ket{0_10_20_30_4}$ as an intermediate state yields a next-nearest neighbor pseudospin exchange interaction $\sigma^+_1 \sigma_2^- + \hc$ (indicate by flipped spins in blue), processes of type (B) induce multi-spin interactions as the energy of intermediate states depends on the configuration of R and L spins and does not average out under summing (symmetry-equivalent) exchange pathways.}
\end{figure}

The effective coupling $J_\mathrm{eff}$ as a function $\theta$ is shown for the relevant range of $\pi/4 \leq \theta \leq 3 \pi/4$ in Fig.~\ref{fig:2nd-order-ising}(b).
Notably, $J_\mathrm{eff}$ changes signs at $\pi/2$ from antiferromagnetic ($J_\mathrm{eff} > 0$) to ferromagnetic ($J_\mathrm{eff} < 0$), concomitant with the sign change of the dipolar exchange coupling $J_1$. Hence there are two distinct outcomes from degeneracy lifting:
\begin{enumerate}
	\item For $\theta > \pi/2$, a partial \emph{ferromagnetic phase} is stabilized which exhibits a spontaneous magnetization $|m^z| = 1/3$ due to $1/3$ of the local moments (here: $C$ sublattice spins) ordering ferromagnetically and the remaining $2/3$ of moments (on $A,B$ sublattices) being in a $S^z=0$ state. Going to higher order in perturbation theory (see also below) will yield additional transverse exchange interactions which will renormalize the magnetization and other observables of interest.
	\item For $\theta < \pi /2$, the effective longitudinal (Ising) coupling between the pseudospins is antiferromagnetic, $J_\mathrm{eff} < 0$, so that the ground state of the system still possesses a macroscopic ground-state degeneracy of $2^{N_C/3} = 2^{N/9}$ due to the intrinsic (Ising) frustration of the triangular superlattice of the $C$-sublattice spins.
\end{enumerate}

While 1.~results in a unique ground state, with higher-order corrections assumed to only renormalize observables and dynamics, higher-order perturbation theory is required for 2.~in order to determine how the degeneracy of the frustrated effective Ising model is lifted.

Expanding on the second scenario, we note that, due to the $S=1$ nature of the local wavefunctions, flipping a pseudospin $\ket{\uparrow} \mapsto \ket{\downarrow}$ in general requires $S^+$ or $S^-$ acting \emph{twice} on the underlying local moment.
We therefore find that transverse exchange processes are only induced at fourth order in $\lambda$.
In the following, we consider the six-spin cluster shown in Fig.~\ref{fig:exge-4th-order}, with spins $1$ and $2$ being $C$-sublattice pseudospin degrees of freedom, and $3,4$ denoting $A,B$-sublattice sites in a $\ket{0}$-configuration.
A pseudospin-flip $\ket{\uparrow_1 \downarrow_2} \mapsto \ket{\downarrow_1 \uparrow_2}$ can occur either (A) via an intermediate state $\ket{0_1 0_2 0_3 0_4}$ (and further appropriate intermediate states to reach that state), or (B) via the intermediate states $\ket{0_1 0_2 \downarrow_3 \uparrow_4}$ or $\ket{0_1 0_2 \uparrow_3 \downarrow_4}$ (and appropriate completion of the fourth-order process), as illustrated in Fig.~\ref{fig:exge-4th-order}.

Summing up all pathways contributing to process (A), we find that the resulting matrix element is independent of the configuration of the two adjacent pseudospins (dubbed ``L'' and ``R''), and furthermore that $\braket{\uparrow_1 \downarrow_2 | h^{(4,i)}_{12} | \uparrow_1 \downarrow_2} = \braket{\uparrow_1 \downarrow_2 | h^{(4,i)}_{12} | \downarrow_1 \uparrow_2} <0$, consistent with an effective XXZ-type nearest-neighbour (for the pseudospins) Hamiltonian with \emph{ferromagnetic} transverse exchange.
On the other hand, matrix elements for processes of type (B) depend on the configuration of the L- and R-spins through the energy of the intermediate states and discriminate between configurations of parallel/anti-parallel L and R-spins, suggesting a contribution to the effective Hamiltonian of the form $(\alpha + \beta \sigma^z_L \sigma^z_R)\sigma^+_1 \sigma^-_2 + \hc$. 
Beyond second order perturbation, higher-order processes are expected to generate multi-spin interactions which, on the triangular lattice, could stabilize a spin-liquid ground state \cite{misguich98,motru05}.
An exhaustive enumeration of those processes, and determining the thereby induced effective exchange couplings is left for future work.

Instead, we remark that if ferromagnetic transverse exchange couplings are the dominant contribution beyond second-order-perturbation theory, the system appears poised to show supersolid order of the $C$-sublattice spins.
In case further (frustrated) multi-spin exchange couplings are of importance, one may speculate that the $C$-sublattice spins are driven into a disordered short-range correlated phase, while the remaining 2/3 background spins are in a unique (nematically) ordered state, bearing similarities to the ``partial quantum disorder'' scenario developed in Refs.~\onlinecite{seif19,gonz19}.  

\section{Spectral signatures} \label{sec:spectr}

\begin{figure*}
	\centering
	\includegraphics[width=\textwidth]{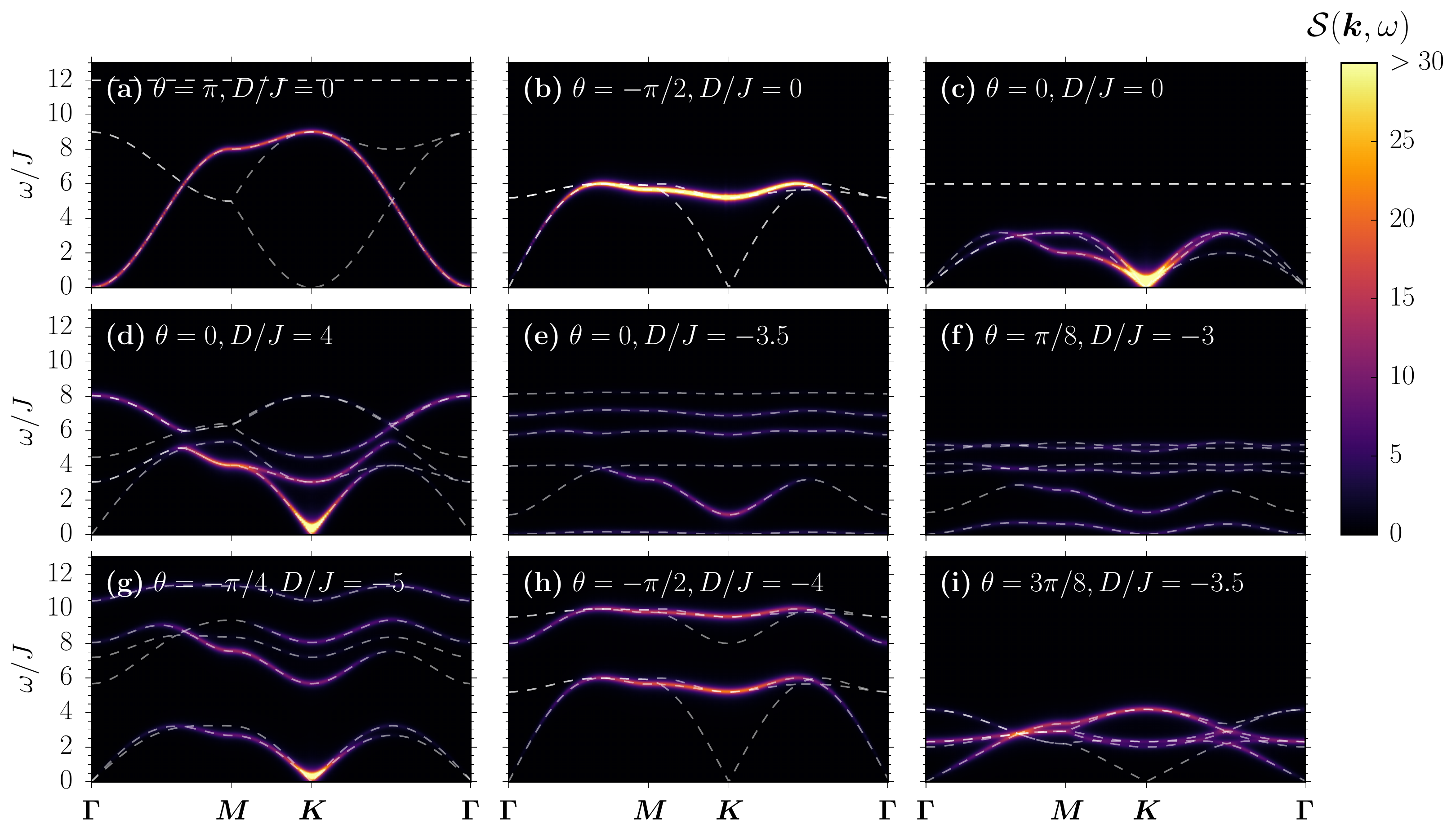}
	\caption{\label{fig:dssLambdaOne}Spectra (white dashed lines) and dynamical spin-structure factor (color-coded) for various parameter regimes in the model with isotropic exchange ($\lambda = 1$) plotted along a path of high-symmetry lines in the full (triangular lattice) Brillouin zone \cite{Note3}. These are the excitation spectra calculated for a mean field ground state (a) in the ferromagnet due to the ferromagnetic Heisenberg model, (b) in the ferroquadrupolar phase due to the ``ferro'' quadrupolar model, (c) in the 120$^\circ$ antiferromagnet due to the antiferromagnetic Heisenberg model, (d) in the 120$^\circ$ antiferromagnet due to the antiferromagnetic Heisenberg model with easy-plane anisotropy, (e) in the zAFM2 phase due to the antiferromagnetic Heisenberg model with easy-axis anisotropy, (f) in the zAFM3 phase due to the ``antiferro'' bilinear-biquadratic model with easy-axis anisotropy, (g) in the zAFM$_{{\rm deg}\,U(1)}$ (``supersolid 1'') phase due to the ``antiferro'' bilinear- ``ferro'' biquadratic model with easy-axis anisotropy, (h) in the ferroquadrupolar phase due to the ``ferro'' quadrupolar model and easy-axis anisotropy, (i) in the antiferroquadrupolar-umbrella phase due to ``antiferro'' bilinear-biquadratic couplings with easy-axis anisotropy.}
\end{figure*}

In this section, we present some representative flavor-wave band structures and corresponding dynamical spin structure factors which reflect the dynamics of excitations in various phase in our model. 

\subsection{General remarks}

The flavor-wave spectra follow straightforwardly from diagonalizing $\Sigma H(\bvec k)$ with $\Sigma = \sigma^z \otimes \mathds{1}_{6 \times 6}$ and $H(\bvec k)$ given in \eqref{eq:H2biq}.
In order to make contact with potential signatures from experiments that are able to resolve spin-spin correlations (in particular neutron scattering), we compute the dynamical spin-structure factor $\mathcal{S}(\bm{k},\omega) = \sum_{\alpha} \mathcal{S}^{\alpha \alpha}(\bm{k},\omega)$ where
\begin{equation} \label{eq:def_structfact}
  \mathcal{S}^{\alpha \beta} (\bvec k, \omega) = \frac{1}{N}\sum_{i,j} \eu^{\iu \bvec k \cdot \left(  \bvec R_i - \bvec R_j \right)} \frac{1}{2 \pi} \int_{-\infty}^\infty \eu^{\iu \omega t} \langle S_i^\alpha S_j^\beta \rangle \du t
\end{equation}
at various points in parameter space, where the wavevectors $\bm{k}$ are elements of the full (triangular lattice) Brillouin zone (BZ) \footnote{Note that working with a $\sqrt{3} \times \sqrt{3}$ unit cell results in backfolding of the linear flavor-wave bands (such that the $\bm{K}$-point of the full BZ is mapped to the $\bm{\Gamma}'$-point of the reduced Brillouin zone associated with the $\sqrt{3} \times \sqrt{3}$ unit cell. The spectra shown are plotted over the full BZ and thus feature identical dispersions at $\bm{\Gamma}$ and $\bm{K}$.}.
For computational details we refer the reader to Appendix~\ref{sec:sfact_detail}.

Note that in flavor-wave theory, the spectrum on top of purely dipolar ordered ground states (i.e.\ $\theta = 0,\pi$ and $D=0$) features flat bands without any weight in the dynamical spin structure factor \cite{batista14}.
The absence of dispersion in these bands is due to a selection rule according to which the purely dipolar Hamiltonian can only have off-diagonal matrix elements among states $\Delta S^z = \pm 1$,
and the vanishing spectral weight results from the fact that the quasiparticle bands can be labelled by $\Delta S^z = \pm 1$ or $\pm 2$, with the latter having vanishing overlap with the state $S^+ \ket{\mathrm{GS}}$ in the DSF.

On the other hand, for any $\theta\neq 0, \pi$ or $D\neq 0$, dipolar and quadrupolar ground-state wavefunctions and/or excited states hybridize such that the selection rule is no longer applicable, and previous ``hidden'' bands can disperse and acquire weight in the spin-structure factor.
These considerations again demonstrate the necessity of using flavor-wave theory as a unified framework for studying excitations of $S>1/2$ magnets, if higher-rank spin-spin interactions (or anisotropies) are thought to be of importance.

\subsection{Results for isotropic exchange, $\lambda = 1$}

\begin{figure*}[htb]
	\centering
	\includegraphics[width=\textwidth]{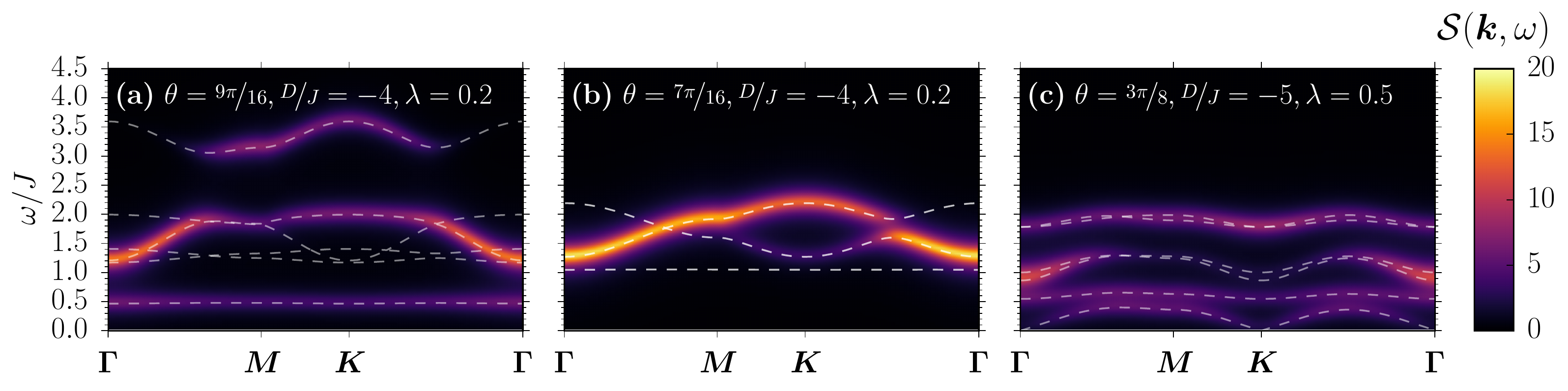}
	\caption{\label{fig:dssLambdaSmall}Spectra (white dashed lines) and dynamical spin-structure factor (color-coded) for three phases stabilized by exchange anisotropies $\lambda < 1$, plotted along a path of high-symmetry lines in the full (triangular lattice) Brillouin zone. Note the different normalization of the colormap compared to Fig.~\ref{fig:dssLambdaOne}, and different scale of the vertical axis. These are the excitation spectra calculated for a mean field ground state (a) in the uuQ phase due to ferro-bilinear and antiferro-biquadratic interactions with easy-axis anisotropy and close the the Ising exchange limit, (b) in the Qud phase due to antiferro-bilinear-biquadratic interactions with easy-axis anisotropy and close the the Ising exchange limit, (c) in the tQud phase due to antiferro-bilinear-biquadratic interactions with easy-axis anisotropy and intermediate transverse exchange anisotropy.}
\end{figure*}

We present the spectra (dashed lines) as well as dynamical spin structure factors (DSS) for various parameter sets in the isotropic model with $\lambda = 1$ in Fig.~\ref{fig:dssLambdaOne}.
Here, panels (a), (b) and (c) depict the spectra and DSS in the absence of any anisotropies, $D=0$ and $\lambda = 1$.
The ferromagnetic phase (a) clearly features a quadratic gap-closing at the $\Gamma$-point, while in the antiferromagnetic case (c) (with 120$^\circ$ N\'eel order) the DSS diverges as one approaches the order-wavevector $K$ (with a linear gap closing corresponding to two degenerate Goldstone modes).
Both phases features ``hidden'' bands due to the aforementioned selection rule.
In the ferroquadrupolar phase (b), one finds linearly dispersing Goldstone modes (at $\Gamma = 0$) with zero spectral weight in the DSS due to their quadrupolar character.
In (d), which corresponds to the easy-plane 120$^\circ$ antiferromagnet (with reduced spin lengths $|\braket{\vec S}|^2 < 1$), one notices that one of the Goldstone modes of (c) has become gapped, and further that the flat ``hidden'' bands have become dispersive and acquired spectral weight around the $\Gamma$ point.

The spectrum and DSS in zAFM2 is shown in panel (e), and exhibits remarkably (quasi-)flat bands, also at the lowest energies, with a linearly closing gap at the $\bm{K}$-point due to the spontaneously broken symmetry of $\Uone$ in-plane rotations.
As one decreases $D/J$ further, the entire band eventually hits $\omega = 0$, signalling the onset of a macroscopic degeneracy found at $\theta = 0$ and sufficiently small $D/J$.
Similarly, in panel (f) we show the spectrum in the zAFM3 ground state. Note that for both of these phases, spin lengths are reduced, indicating the mixed character of quadrupolar and dipolar local wavefunctions, resulting in a strong hybridization of magnons and quadrupolar waves and thus reduced weights in the dynamical spin structure factor.

A representative spectrum for the supersolid phase is shown in Fig.~\ref{fig:dssLambdaOne}(g).
Note that at low energies, two gapless modes (with a linear dispersion) at $\bm{K}$ are visible, namely a Goldstone mode corresponding to the spontaneous breaking of the in-plane $\Uone$ spin rotation symmetry, as well as a ``degeneracy mode'' due to the accidental $\Uone$ degeneracy of ground states in the effective XXZ model \cite{murthy97}.
As discussed in Sec.~\ref{sec:region-A}, this accidental degeneracy is lifted via the order-by-disorder mechanism, and as discussed in Ref.~\onlinecite{murthy97}, the selected state will have a maximal number of independent gapless modes, while generically, for other states in the accidental $\Uone$-manifold, the physical Goldstone mode and degeneracy mode mix and do not constitute independent modes.
The appearance of two linearly gap-closing modes in Fig.~\ref{fig:dssLambdaOne} thus confirms the validity of our ground-state selection protocol discussed in Sec.~\ref{sec:region-A}.

Further, in panels (h) and (i) we show flavor-wave spectra in the easy-axis ferroquadrupolar phase (where one Goldstone mode now has become gapped and yields a high-energy band visible in the dynamical spin structure factor), and in the AFQ-U phase, where the directors form an umbrella-type configuration.

\subsection{Results for XXZ-exchange, $\lambda < 1$}

As discussed in Sec.~\ref{sec:exact-Ising-lim}, for sufficiently strong anisotropies $\lambda \ll 1$ novel ordered phases can be realized which are adiabatically connected to corresponding phases in the exactly solvable generalized $S=1$ Ising (Blume-Capel) model on the triangular lattice.

\begin{figure}[htb]
	\centering
	\includegraphics[width=\columnwidth]{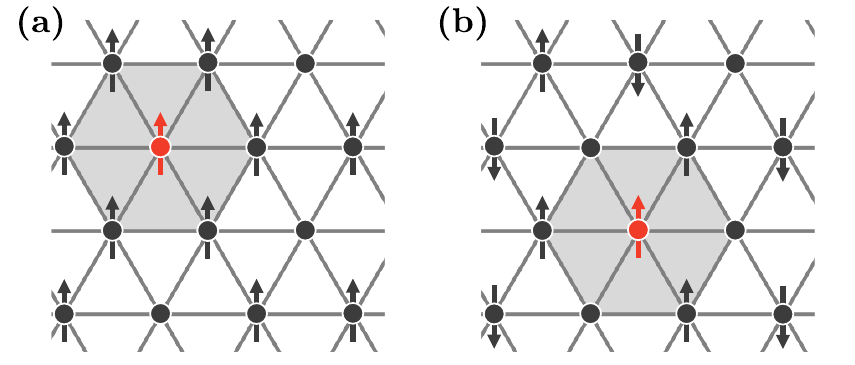}
	\caption{\label{fig:ising-singleExc}(a) Localized lowest-energy excitation in the ``$\uparrow\uparrow0$''-phase in the $S=1$ Ising (Blume-Capel) model, adiabatically connected to the nearly flat band at $\lambda >0$ shown in Fig.~\ref{fig:dssLambdaOne}(a).}
\end{figure}

In Fig.~\ref{fig:dssLambdaSmall}, we present exemplary spectra and dynamical spin structure factors for some of these phases.
A particularly striking feature in panels (a) and (b) is the emergence of isolated almost perfectly flat bands as lowest energy excitations.
Note that while not visible to the naked eye, we find that, in the parameter regimes shown, these bands have a small (but finite!) bandwidth of $\Delta E \sim 10^{-2}J$ \footnote{We have verified that this is not due to the propagation of numerical errors, occurring in the numerical optimization of the ground-state parameters in Eq.~\eqref{eq:dvec-angles} by (1) varying the working precision of our numerical search algorithm and (2) constructing analytical matrices $U$ in \eqref{eq:umat} and using them to construct the dynamical matrix $H(\bm{k})$ which is then Bogoliubov-diagonalized numerically.}.
Tuning $\lambda \to 0$, we find that these bands are adiabatically connected to {\em fully localized} excitations in the exactly solvable $\lambda=0$-model:
In the case of $\theta > \pi/2$, shown in Fig.~\ref{fig:dssLambdaSmall}(a) (small ferromagnetic dipolar interactions), when the $\uparrow\uparrow0$ ground state is stabilized, the low-lying flat band evolves from the localized excitation corresponding to exciting $S^z=0 \to S^z= + 1$ on a single site, resulting in six defect triangles and a total energy of $\Delta E /J= D/J + 6 (\cos \theta + \sin \theta)$, as illustrated in Fig.~\ref{fig:ising-singleExc}(a).
This excitation carries $\Delta S^z = \pm 1$ and thus the band at $\lambda>0$ has finite weight in the dynamical spin structure factors.
A qualitative argument for the flatness of this band consists in noting that to first order in perturbation theory in $\lambda$, the transverse pieces of the Hamiltonian do not give rise to a hopping term of this excitation.
Instead, a dispersion is only induced by a twofold application of dipolar transverse exchange $S^+_i S^-_j + \hc$, i.e.\ appear at order $\lambda^2$.
However, we emphasize the qualitative nature of this argument, as the quasiparticles of the model at any finite $\lambda$ are not spin flips but rather coherent superpositions of the Schwinger Bosons in Eq.~\eqref{eq:s_schwinger_bosons} and thus no longer carry well-defined $\Delta S^z$-quantum numbers.
 
Similarly, for the phase ``$\uparrow\downarrow\! 0$'', stabilized for $\theta < \pi/2$ and small $\lambda$, we find that the flat band found for small $\lambda$, visible in Fig.~\ref{fig:dssLambdaSmall}, evolves from the spin-flip excitation $S^z = \pm 1 \to S^z = \mp 1$ on a single site (as illustrated in Fig.~\ref{fig:ising-singleExc}(b)), with energy difference $\Delta E = 3 J\cos \theta$. The fact that this excitation at $\lambda = 0$ excitation has $\Delta S^z = 2$ renders it invisible (zero weight) in the DSS.
Arguing perturbatively, it becomes clear that a hopping term for this excitation can only be induced through a higher-order process involving the twofold application of $(S^+_i S^-_j)^2 + \hc$, and in turn the hopping matrix element scales with $\lambda^4$ and involves higher-energy intermediate excitations.
We again note the qualitative nature of this argument, finding that the band is slightly dispersive in linear flavor-wave theory.

Further, we show the spectrum and DSS in the tQud phase in which the $z$-components of the spins on the three sublattices are analogous to the ``$\uparrow\downarrow\! 0$''-state, but with an additional spontaneous breaking of the in-plane $\Uone$ spin rotation symmetry as the dipolar moments tilt away from the $\hat{z}$-axis, and on the sublattice $\langle \vec{S} \rangle = 0$ the quadrupole tensor $Q^{\alpha \beta}$ breaks the rotational symmetry.
Consequently, one observes a Goldstone mode at $\bm{K}$ which we can attribute to be of primarily quadrupolar character due to the vanishing spectral weight in $\mathcal{S}(\bm k,\omega)$.

\section{Conclusion} \label{sec:conclusio}

\subsection{Summary}

In this work, we have revisited the spin-1 bilinear-biquadratic model on the triangular lattice with a single-ion anisotropy.
We have shown that adding an easy-axis exchange anisotropy, corresponding to an XXZ-interaction in the bilinear sector, allows one to discover new phases and study how unconventional magnetic orders emerge from the (frustrated) generalized $S=1$ Ising (Blume-Capel) model on the triangular lattice as one includes transverse exchange interactions.

In particular, we were able to use perturbation theory  in the transverse exchange coupling on top of the degenerate antiferromagnetic Ising ground-state manifold to argue for the existence of a supersolid phase, constituting a complimentary approach to the previously employed limit of large single-ion anisotropy.
Moreover, we have found that the in the Ising limit, the model's phase diagram features an additional macroscopically degenerate phase with $S^z=0$ on two sublattices and a degenerate $S^z= \pm 1$ on the remaining sublattice, with the degeneracy found to persist for finite transverse exchange in mean-field theory.
Using perturbation theory we have found that for ferromagnetic transverse exchange couplings the degeneracy is lifted in favor of a $1/3$-polarized ferromagnet, while for antiferromagnetic transverse exchange coupling higher-order perturbation theory is required, likely inducing multi-spin exchange couplings which may drive the system to either a quantum spin liquid or to a quantum-disordered short-range correlated phase.

Having implemented a protocol to (numerically) perform a linear-flavor wave expansion on top of \emph{any} variational ground state of the system, we have calculated the first-order quantum corrections to the moment of the order parameter.
We find that these corrections are particularly strong at points of enhanced symmetry as well as in regions with multiple competing phases.

We have further shown that in some of the Ising-type phases realized for strong XXZ anisotropies, the flavor-wave spectra exhibit strikingly flat bands which can be related to localized excitations in the Ising case.

\subsection{Outlook}

Our work is expected to be applicable to spin-1 triangular lattice antiferromagnets studied experimentally in recent years, such as FeGa$_2$S$_4$ \cite{gura21} or Ba$_3$NiNb$_2$O$_9$, which appears to possess $120^\circ$ order with an easy-plane exchange anisotropy \cite{hwang12,lu18} and exhibits an unusual ``oblique'' phase at intermediate applied field strengths.
To our knowledge, the role of biquadratic exchange interactions in these systems has not been elucidated yet. 
Moreover, recent numerical and theoretical studies \cite{kart20,ni21} have suggested that sizeable biquadratic exchange interactions are of relevance to ordered states in some systems in the recently discovered and experimentally studied class of van-der-Waals magnets \cite{novo19,ralph19}, with additional exchange anisotropies likely present \cite{kart20}.

On the theoretical front, our work opens up several avenues for further study: It would be of interest to obtain numerical results that verify the existence of supersolid phases in the full $S=1$ model, and their nature for dominantly antiferromagnet exchange and large easy-axis single-ion anisotropies (region $\mathcal{A}$).
Similarly, the perturbative degeneracy lifting mechanism for the macroscopically degenerate mixed dipolar-quadrupolar phase found for $\theta \simeq \pi/2$ (region $\mathcal{B}$) calls for further study, in particular for the case of $\theta < \pi/2$ where we find that only higher-order processes could lift the degeneracy.
For accurate modelling of microscopic materials, additional anisotropy terms are conceivable, with the study of magnetic-field induced phases as a useful constraint on allowed parameter sets.

It would also be useful to determine what resonant (inelastic) x-ray scattering (RXS) signals are produced by the states in our phase diagrams as RXS is better tailored than neutron scattering to image quadrupolar phases and their excitations.\cite{savary2015} Finally, it would be interesting to systematically explore under what conditions (higher-spin) Ising ground states with complex ordering patterns provide a platform for obtaining (nearly) flat magnon (or quadrupole-wave) bands, and to what extent this allows for the engineering of topologically non-trivial bands \cite{chis15,ma20}.

\acknowledgements

We thank Radu Coldea for stimulating discussions. This work was funded by the European Research Council (ERC) under the European Union's Horizon 2020 research and innovation program (Grant agreement No.\ 853116, acronym TRANSPORT).

\clearpage
\appendix

\section{Technical details on linear flavor wave theory}

\subsection{Construction of transformation to canonical basis} \label{sec:def_trafo}

Constructing an appropriate transformation $U$ in \eqref{eq:umat} amounts to finding two vectors $\vec u$ and $\vec v$ in Eq.~\eqref{eq:umat} which, together with $\uvec d$, will form an orthonormal basis of $\mathbb{C}^3$ with the standard scalar product $\langle u, v \rangle = {\uvec u}^\dagger {\uvec v}$.
To this end, we pick a trial vector, say $(1,1,1)^\top$, and perform a Gram-Schmidt-orthogonalization step (recall that $\uvec d^\dagger \uvec d =1$)
\begin{equation}
	\uvec u^{(0)} \mapsto (1,1,1)^\top- ({\uvec d}^\dagger (1,1,1)^\top)\uvec d \quad \text{and} \ \uvec u^{(1)} = \uvec u^{(0)} / || \uvec u^{(0)} ||.
\end{equation}
For the second vector $\uvec v$, one could proceed with another trial vector analogously, but instead we note that for any $\uvec u$ we have
\begin{equation}
	\uvec u^\dagger (\uvec u \times \uvec d)^\ast = \uvec d^\dagger (\uvec u \times \uvec d)^\ast = 0 \quad \text{and} \quad \left( (\uvec u \times \uvec d)^\ast \right)^\dagger  (\uvec u \times \uvec d)^\ast = 1
\end{equation}
so that we may take $\uvec v = \left(\uvec u \times \uvec d\right)^\ast$.
Note that this procedure fails if $\uvec d^\dagger \equiv (1,1,1)^\top$, in this case, a different trial state $(1,0,-1)^\top$ may be picked. 

\subsection{Computation of dynamical structure factor} \label{sec:sfact_detail} 

We take the dynamic structure factor to be given by Eq.~\eqref{eq:def_structfact}.
We can factorize the spatial Fourier transformation to write
\begin{align}
  1/N\sum_{i,j} \eu^{\iu \bvec k \cdot \left(  \bvec R_i - \bvec R_j \right)} \langle S^\alpha_i S^\beta _j \rangle = &1/N \langle \left( \sum_i S_i^\alpha \eu^{\iu \bvec k \cdot \bvec R_i} \right) \nonumber\\ &\times\left( \sum_j S_j^\beta \eu^{-\iu \bvec k \cdot \bvec R_j} \right) \rangle.
\end{align}
Importantly, we can write the positions of the spins as $\bvec R_i = \bvec r_i + \bvec \delta_{s(i)}$, where $\bvec r_i$ are primitive lattice vectors connecting $\sqrt{3} \times \sqrt{3}$ unit cells and $\bvec \delta_{s(i)}$ is the position of the spin at site $i$ within the unit cell (i.e. the sublattice of site $i$). We choose the convention that $\bvec{\delta}_A = 0$, and thus $\bvec{\delta}_B = (-1/2,\sqrt{3}/2)$, $\bvec{\delta}_C = (1/2,\sqrt{3}/2)$.
Thus
\begin{equation}
  \frac{1}{\sqrt{N}}\sum_i S_i^\alpha \eu^{\iu \bvec k \cdot \bvec R_i} = \frac{1}{\sqrt{3}}\sum_s \underbrace{\frac{1}{\sqrt{N_\mathrm{uc}}}\sum_{j \in \text{unit cells}}  S_{j,s}^\alpha \eu^{\iu \bvec k \cdot r_j}}_{=S^\alpha_{s,-\bvec k}} \eu^{\iu \bvec k \cdot \bvec \delta_s}
\end{equation}
Thus we have for the sublattice-resolved structure factor
\begin{equation}
  \mathcal{S}^{\alpha \beta}_{s,s'}(\bvec k,\omega) = \langle S^\alpha_s (-\bvec k,- \omega)S^\beta_{s'}(\bvec k, \omega) \rangle \eu^{\iu \bvec k \cdot (\bvec\delta_s - \bvec\delta_{s'})}
\end{equation}
We express the spin operators in terms of the three Schwinger bosons (here the spinor $\uvec b = (b_x,b_y,b_z)^\top$) and perform a change to the canonical basis
\begin{equation}
  S_{i,s}^\alpha = \uvec{b}^\dagger_{i,s} (-\iu \epsilon^\alpha ) \uvec{b}_{i,s} = {\uvec{b}'}_{i,s}^\dagger U_s (- \iu \epsilon^\alpha ) U_s^\dagger \uvec{b}'_{i,s}
\end{equation}
where $(\epsilon^\alpha)_{\beta \gamma} = \epsilon^{\alpha \beta \gamma}$.
We then condense the $b_z'$-boson, expand in $1/M$ as usual and Fourier-transform, yielding
\begin{align}
  S^\alpha_{\bvec k,s} = M \times \mathrm{const.} + \sqrt{M}\big( &m^\alpha_{s,\mu}(\bvec k) \psi_{s,\mu}(\bvec k) \nonumber\\ &+ (m^\alpha_{s,\mu}(\bvec k))^\ast \psi_{s,\mu}^\dagger(-\bvec k) \big)
\end{align}
where $\psi_{s,1} = b_{x,s}$ and $\psi_{s,2} = b_{y,s}$, and $m^\alpha_{s,\mu}$ are some coefficients (note that summation over repeated indices is implied unless otherwise noted)
The dynamical contributions to the static structure factor are obtained at order $M$ and read (after setting $M=1$)
\begin{align}
&{\mathcal{S}}^{\alpha \beta}(\bvec k, \omega)_{s,s'} = \sum_n \braket{\emptyset|{S}^\alpha_{s,-\bvec k } | n } \braket{n | \underbrace{S^\beta_{s',\bvec k }}_{\sim (m_{s'})_X^\beta \psi_X} | \emptyset} \delta(\omega- (\omega_n - \omega_0)) \nonumber\\
 &= \sum_n (m_s)^\dagger_{\alpha X} (T^\dagger)_{\tau X} \braket{\emptyset| \gamma_\tau^\dagger| n} \nonumber\\ &\quad \times \braket{n|\gamma_\lambda|\emptyset} T_{Y\lambda} (m_{s'})_{Y \beta} \delta(\omega- (\omega_n - \omega_0)),
\end{align}
where $\ket{\emptyset}$ denotes the vacuum of the eigenmodes (which in general is \emph{not} equivalent to $\ket{S^z=0}$).
The matrix $T$ is the $12 \times 12$ Bogoliubov rotation matrix, and we have introduce a $12$-component spinor $\Psi = (\psi_{A,1}, \dots, \psi_{B,2}, \psi_{A,1}^\dagger, \dots \psi_{B,2})^\top$ which is indexed by composite indices $X$ (and similar for the coefficients $(m_s)_{\alpha X}$.

In practice, we use \texttt{Mathematica} to expand the spin operators on the three sublattices and thus get three distinct $m$-matrices $m_s$ for $s=A,B,C$ sublattice sites.
Matrix elements $\braket{n|\gamma_\lambda|\emptyset}$ are only finite if $\gamma$ is a creation operator, which is determined from the conventions employed in the Bogoliubov transformation. Here, we have chosen the last 6 elements of $\psi$ (and thus of $\gamma$) to be creators, and thus $\Sigma^z$, $\braket{\emptyset| \gamma_\tau^\dagger| n} \braket{n|\gamma_\lambda|\emptyset} = \frac{1}{2} (\mathds{1}-\Sigma^z)_{\tau\lambda} = (0,\dots,0,1,\dots,1)$.
We thus find
\begin{align}
  &{\mathcal{S}}^{\alpha \beta}_{s,s'}(\bvec k, \omega) = \sum_\lambda ({m_s}^\dagger)_{\alpha X} (T^\dagger)_{\tau X} \frac{1}{2} (\mathds{1}-\Sigma^z)_{\tau\lambda} T_{Y\lambda} {m_{s'}}_{Y \beta} \nonumber\\
  &\quad \quad \times\delta(\omega- \omega_\lambda(\bvec k)) \eu^{\iu \bvec k \cdot (\bvec\delta_s - \bvec\delta_{s'})} \nonumber\\
  &= \left[{m_s}^\dagger T^\ast \left[ \frac{1}{2} (\mathds{1}-\Sigma^z) \right] \underline{\diag[ \delta(\omega- \omega_\lambda(\bvec k) )]} T^\top (m_{s'}) \right]^{\alpha \beta} \nonumber\\
  & \quad\quad \times \eu^{\iu \bvec k \cdot (\bvec\delta_s - \bvec\delta_{s'})},
\end{align}
where in the last equality we have rewritten the expression in a matrix form which can be straightforwardly implemented in mathematica, if the $\delta$-function is viewed as a $12 \times 12$ diagonal matrix, indicated by the underline.
The full structure factor is then $\sum_{s,s'} \mathcal{S}^{\alpha \beta}_{s,s'} (\bvec k, \omega)$.

\section{First-order correction to local moments}

As elaborated in the main text, to quantify the renormalization of the amplitude of the local order parameter it is sufficient to compute the norm $\langle \uvec{b}^\dagger \rangle \langle \uvec b \rangle \equiv M - \langle b_x'^\dagger b_x' + b_y'^\dagger b_y' \rangle$, from which the amplitude $m$ of any order parameter is obtained as
\begin{align}
	m = \frac{1}{N} \sum_i &\langle \uvec{b}_i^\dagger \rangle  \langle \uvec b_i \rangle  \equiv M - \Delta M  \nonumber\\ & \text{where} \quad \Delta M = \frac{1}{N} \sum_i \langle b_{i,x}'^\dagger b_{i,x}' \rangle + \langle b_{i,y}'^\dagger b_{i,y}' \rangle
\end{align}
and we seek to compute $\Delta M$.
Fourier-transforming, we can write
\begin{align} \label{eq:def-delta}
	\Delta M &= \frac{1}{N_\mathrm{uc}} \sum_{\bvec{k}} \langle {\psi}_\bvec{k}^\dagger \begin{pmatrix}
		\mathds{1} & 0 \\
		0 & 0
	\end{pmatrix} {\psi}_\bvec{k} \rangle \\ &= \frac{1}{N_\mathrm{uc}} \sum_{\bvec{k}} \frac{1}{4} \tr  \left[ {T}(\bvec{k})^\dagger (\mathds{1}+\Sigma^z)
	 {T}(\bvec{k}) (\mathds{1}-\Sigma^z) \right],
\end{align}
where $\Psi_{\bvec k}$ denotes the spinor introduced in \eqref{eq:H2biq}, and the second equality follows from using the definition of the Bogoliubov rotation ${\Psi}_\bvec{k} = {T}(\bm{k}) {\gamma}_\bvec{k}$ and evaluating the expectation values of the normal modes (at $T=0$ temperature). 

We note that in several phases the dispersion features gapless points associated with Goldstone modes and/or (accidental) degeneracies.
At these points, the matrix ${H}(\bvec{k})$ is only positive semi-definite (rather than positive definite), making the Cholesky-decomposition in Colpa's algorithm ill-defined \cite{colpa}.
We emphasize that, in $d=2$ dimensions and at zero temperature $T=0$, Goldstone modes do not destroy order as guaranteed by the Mermin-Wagner-Theorem \cite{merminwag} and thus $\Delta$ needs to be regular (i.e. the momentum-space integration does not suffer a infrared divergence) \footnote{Note however, that $\Delta > 1$, signalling that \emph{quantum} fluctuations destabilize the magnetic order}.
 
 With this in mind, we proceed by evaluating \eqref{eq:def-delta} on momentum-space grids of $N \times N$ unit cells where $N = 32,48,64$ with the $\Gamma$-point $k=0$ removed (recall that we work in a reduced $\sqrt{3} \times \sqrt{3}$ Brillouin zone so that the crystallographic $K$-points are mapped onto the $\Gamma$-point).
 We then perform a finite-size extrapolation $\Delta M(N) =  A \frac{1}{N} + (\Delta M)_\infty$ with $A$ some scaling factor to obtain the infinite-system result $(\Delta M)_\infty \equiv \Delta M$.
 As a non-trivial cross-check, we find that in the pure (dipolar) Heisenberg limit $\theta = 0$, $D = 0$ and $\lambda =1$ the order parameter correction $\Delta M = 0.261$ which precisely matches the result obtained in previous (conventional) linear spin-wave computations \cite{joli89,chubu94}.
 (We emphasize that the comparison of $\Delta M$ as obtained in linear flavor-wave theory and the conventional spin-wave theory is justified as for $\theta = 0$ and $D=0$, the Hamiltonian for the flavor bosons splits into a magnon ($S=1$) and a quadrupolar ($S=2$) sector, with the latter already being diagonal and thus not requiring a non-trivial Bogoliubov transformation).

\begin{table*}
\caption{\label{tab:states}Overview of states found using the variational mean-field ground-state search, and one corresponding representation in terms of $\uvec{d}$-vectors. Here, $s\in\{A,B,C\} \equiv \{0,1,2,3\}$ and $\umat{R}^\alpha(\phi)$ denotes a $\SO(3)$ rotation matrix about the $\alpha$-axis by the angle $\phi$. If values of the parameters $\phi_s$ and $\eta$ are not specified, they need to be determined variationally.}
\begin{xtabular*}{\textwidth}{c p{0.08\textwidth}p{0.5\textwidth}p{0.28\textwidth}}
		Illustration & Name & Description & $\uvec{d}$-vectors for a ``minimal'' representative configuration  \\ \hline\hline
		\adjustbox{valign=t,margin=\marginv}{\includegraphics[scale=\scalef]{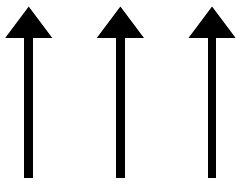}} & FM & Ferromagnet with in/out-of-plane  moments $\vec S$ for $D< 0$ ($D>0$) & $\uvec{d}_s = (0,-\iu \sin (\eta/2),\cos (\eta/2))^\top$ with $\eta = \pi$ for $D <0$, $\eta\in \{0,\pi \}$ for $D>0$   \\
		\hline\adjustbox{valign=t,margin=\marginv} {\includegraphics[scale=\scalef]{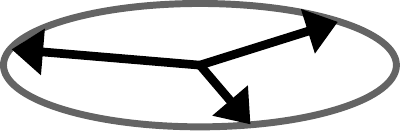}} & $120\,{}^\circ$  AFM & Néel antiferromagnet with in-plane moments for $D>0$ & $\uvec{d}_s = \umat{R}^z(\phi_s) (0,\sin (\eta/2),\cos(\eta/2))^\top$ with $\phi_s = 2 \pi s /3$ \\
		\hline\adjustbox{valign=t,margin=\marginv}{\includegraphics[scale=\scalef]{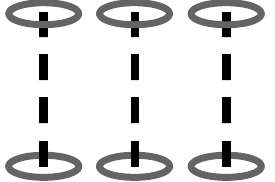}} & FQ & Parallel aligned in/out-of-plane directors $\vec d$ for $D<0$ ($D>0)$ & $\uvec{d}_s = (0,0,1)^\top$ for $D>0$ and $\uvec{d} = (1,0,0)^\top$ for $D < 0$  \\
		\hline\adjustbox{valign=t,margin=\marginv}{\includegraphics[scale=\scalef]{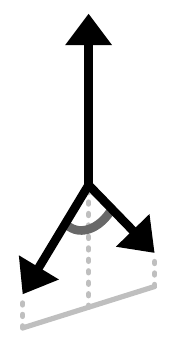}} & zAFM1 & Coplanar easy-axis antiferromagnet with one moment fully polarized $\parallel \hat{z}$, two moments with reduced $|\langle \vec S \rangle|^2 < 1$ and equal angles to first spin, total moment $\langle \vec S_\mathrm{tot} \rangle \parallel \hat{z}$ & $\uvec{d}_s = \umat{R}^x(\phi_s) (-\iu \cos \eta_s, \sin \eta_s,0)^\top$, with $\phi_A = 0, \phi_B = -\phi_C$ and $\eta_B = \eta_C$. \\
		\hline\adjustbox{valign=t,margin=\marginv}{\includegraphics[scale=\scalef]{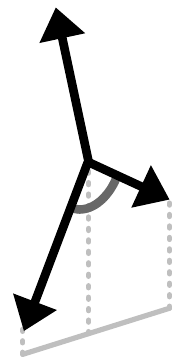}} & zAFM2 & Coplanar easy-axis antiferromagnet similar to zAFM1, but no spin parallel to $\hat{z}$, and inequivalent mutual angles. Two spins dominantly dipolar $\langle \vec S_i \rangle^2 \approx 1$. Total moment $\langle \vec S_\mathrm{tot} \rangle \neq 0$ has both in-and out-of-plane components. & $\uvec{d}_s = \umat{R}^x(\phi_s) (-\iu \cos \eta_s, \sin \eta_s,0)^\top$, with \emph{only} $\phi_C = 4 \pi / 3$, all other parameters unconstrained \\
		\hline\adjustbox{valign=t,margin=\marginv}{\includegraphics[scale=\scalef]{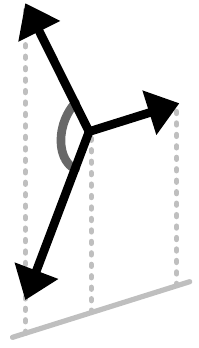}} & zAFM3 & Similar to zAFM1, but rotated: One dipolar moment with strongly reduced length lies in the $\hat{x}\hat{y}$-plane, with remaining two moments symmetrically rotated out of plane (first spin bisects the angle\footnote{Note that this angle is smaller than $ \pi$ for the phase zAFM3 shown in Fig.~\ref{fig:pd_theta_D}, but greater than $ \pi$ for the parameter range displayed in Fig.~\ref{fig:lambdaCut_3pi8}.} between the remaining two moments). Total moment lies in the plane, $\langle \vec S_\mathrm{tot} \rangle \perp \hat{z}$. & $\uvec{d}_s = \umat{R}^x(\phi_s) (-\iu \cos \eta_s, \sin \eta_s,0)^\top$, with $\phi_A = \pi/2$, $\phi_B = -\phi_C$ and $\eta_C = \eta_B$. \\
		\hline\adjustbox{valign=t,margin=\marginv}{\includegraphics[scale=0.5]{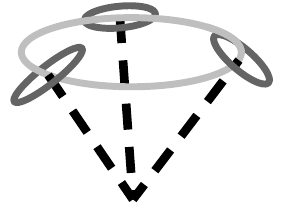}} & AFQ-U & Directors form ``umbrella'' configuration (AFQ-U), with the angle to the common axis (here: $\hat{z}$) varying continuously. & $\uvec{d}_s = \umat{R}^z(\phi_s)\umat{R}^x(\eta)(0,0,1)^\top$ with $\phi_s = 2 \pi s /3$ \\
		\hline\adjustbox{valign=t,margin=\marginv}{\includegraphics[scale=0.5]{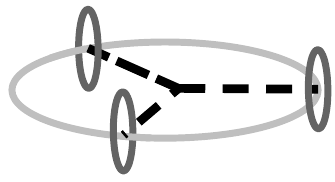}} & AFQ-P & Quadrupolar analogue of $120^\circ$ N\'eel order, directors lie in $\hat{x}{y}$-plane. Relative $2 \pi/3$ angles between directors.
		& $\uvec{d}_s = \umat{R}^z(\phi_s) (1,0,0)^\top$ where $\phi_s = 2\pi s/3$ \\
		\hline\adjustbox{valign=t,margin=\marginv}{\includegraphics[scale=0.55]{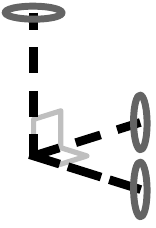}} & AFQ$_{\pi/2}$ & Quadrupolar state with director parallel to $\hat{z}$-axis, two directors in-plane. All mutual angles $\pi/2$.\textsuperscript{\ref{fn:dir}} & $\uvec{d}_A = (1,0,0)^\top$, $\uvec{d}_B = (0,1,0)^\top$ and $\uvec{d}_C = (0,0,1)^\top$ \\
		\hline\adjustbox{valign=t,margin=\marginv}{\includegraphics[scale=0.55]{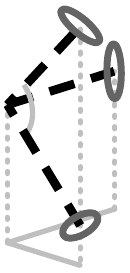}} & AFQ$_\text{xy}$ & One director lies in the $\hat{x}\hat{y}$ plane, the remaining two vectors lie in an orthogonal plane to it (mirror symmetry through $\hat{x}\hat{y}$-plane) & $\uvec{d}_A = (1,0,0)^\top$, $\uvec{d}_B = (0,\cos \phi, \sin \phi)^\top$ and $\uvec{d}_C = (0,\cos \phi,-\sin \phi)^\top$.\\
		\hline\adjustbox{valign=t,margin=\marginv}{\includegraphics[scale=0.55]{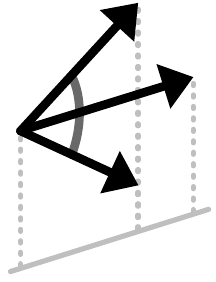}} & Fan$_0$ & Fan-like coplanar ferromagnet, with one moment lying in the plane, and the remaining two moments being reflections of each other along the $\hat{x}$-$\hat{y}$-plane. Note that there is no total moment along $\hat{z}$. & $\uvec{d}_s = \umat{R}^x(\phi_s) (- \iu \cos \eta_s, \sin \eta_s,0)^\top$ with $\phi_A = -\phi_B$, $\phi_C = \pi /2$ and $\eta_A = \eta_B$. \\
		\hline\adjustbox{valign=t,margin=\marginv}{\includegraphics[scale=0.55]{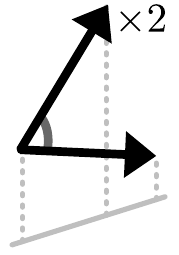}} & Fan$_\mathrm{z}$ & Fan-like coplanar ferromagnet, with moments on two sublattices parallel (with identical ``spin lengths'' $|\langle \vec S \rangle| < 1$, forming a finite angle with the moment on the third sublattice. Note that the common-plane contains the $\hat{z}$-axis, and there is a finite magnetization along $\hat{z}$. & $\uvec{d}_s = \umat{R}^x(\phi_s) (- \iu \cos \eta_s, \sin \eta_s,0)^\top$ with $\phi_A = \phi_B$ and $\eta_A = \eta_B$. \\
		\hline\adjustbox{valign=t,margin=\marginv}{\includegraphics[scale=0.55]{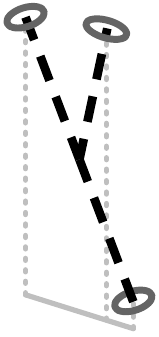}} & AFQ-V & Directors are coplanar. Two directors are parallel with a non-zero angle to the $\hat{z}$-axis. The third director forms a different angle with the $\hat{z}$-axis. & $\uvec{d}_A = (0, \sin \eta,\cos \eta)^\top$ and $\uvec{d}_B = \pm \uvec{d}_C = (0, \sin \phi, \cos \phi)^\top$.
		\end{xtabular*}
\end{table*}
\begin{table*}
\begin{xtabular*}{\textwidth}{c p{0.08\textwidth}p{0.5\textwidth}p{0.28\textwidth}}
		\hline\adjustbox{valign=t,margin=\marginv}{\includegraphics[scale=\scalef]{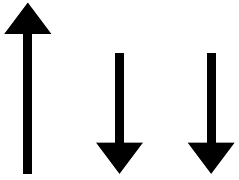}} & zAFM$_\text{deg}$ & Easy-axis antiferromagnet with $d^z = 0$ on all sublattices, described by effective $S=1/2$ XXZ model for $\ket{\uparrow,\downarrow} \equiv \ket{\pm 1}$ pseudospin. Mixed dipolar/quadrupolar character (dipolar moments $\parallel \hat{z}$). Accidental $\Uone$ degeneracy lifted by order-by-disorder. & $\uvec{d}_s = \cos \eta_s (1,\iu,0)^\top \ \sqrt{2} + \sin \eta_s (1,-\iu,0)^\top / \sqrt{2}$ with $\eta_A = 0$, $\eta_B = - \eta_C$  \\
		\hline\adjustbox{valign=t,margin=\marginv}{\includegraphics[scale=\scalef]{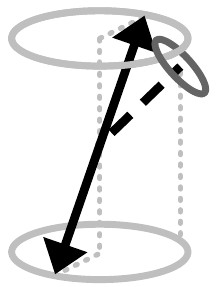}} & tQud (tilted quadru\-polar-up-down) & Dipolar moments pointing in opposite directions on the two sublattices (with $\langle S^z \rangle \gg \langle S^x \rangle, \langle S^y \rangle$), third moment is purely quadrupolar with $\langle \vec S \rangle = 0$.  & \multicolumn{1}{c}{---\footnote{\label{fn:nosimple}These ordering patterns require several independent parameters to be determined variationally, and no ``simple'' parametrization can be given.}}\\
		\hline\adjustbox{valign=t,margin=\marginv}{\includegraphics[scale=\scalef]{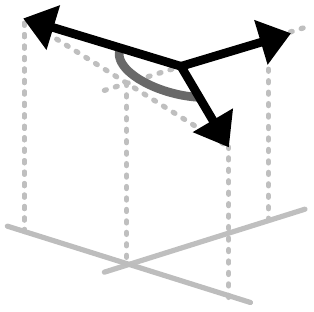}} &  $\kappa_z$-AFM & Coplanar spin configuration with one spin (e.g. on $C$ sublattice) in the $\hat{x}\hat{y}$-plane. Reduced moments $\langle \vec S_A \rangle^2 = \langle \vec S_B \rangle^2 < \langle \vec S_C \rangle^2$. The common plane does not contain the $\hat{z}$ axis such that there is a finite out-of-plane spin chirality $\kappa_z \sim \vec S_A \times \vec S_B + \vec S_B \times \vec S_C + \vec S_C \times \vec S_A$. & \multicolumn{1}{c}{---\footref{fn:nosimple}}
		\end{xtabular*}
\end{table*}


\bibliography{sonetri}

\bibliographystyle{apsrev4-1}

\end{document}